\pdfoutput=1
\documentclass{article}

\usepackage{PRIMEarxiv}
\usepackage{amsmath,amssymb,amsfonts}
\usepackage{graphicx}
\usepackage{textcomp}
\usepackage{xcolor}
\usepackage[utf8]{inputenc}
\usepackage[T1]{fontenc}
\DeclareUnicodeCharacter{2011}{-}
\DeclareUnicodeCharacter{2013}{--}
\DeclareUnicodeCharacter{2018}{`}
\DeclareUnicodeCharacter{2019}{'}
\DeclareUnicodeCharacter{201C}{``}
\DeclareUnicodeCharacter{201D}{''}
\usepackage{booktabs}
\usepackage{array,ragged2e}
\usepackage{algorithm}
\usepackage{algpseudocode}
\usepackage{braket}
\usepackage[font=footnotesize]{subfig}
\usepackage{fancyhdr}
\newcommand{\orcidlink}[1]{}
\begin{document}

\title{Learning Temporal Patterns in Financial Time Series: A Comparative Study of Quantum LSTM and Quantum Reservoir Computing}

\author{%
\textbf{Danyal Maheshwari,
Gerhard Hellstern,
Martin Zaefferer,
Martin Braun, and
Tanja D\"ohler}
}
\date{}
\maketitle 
\begin{center}
\textit{Centre of Finance, DHBW Stuttgart}, Stuttgart, Germany \\
\texttt{danyal.maheshwari@dhbw-stuttgart.de}, \texttt{gerhard.hellstern@dhbw-stuttgart.de} \\
\vspace{0.3em}
\textit{Zentrum f\"ur Digitale Innovationen, DHBW Ravensburg}, Ravensburg, Germany \\
\texttt{zaefferer@dhbw-ravensburg.de} \\
\vspace{0.3em}
\textit{DATEV eG}, N\"urnberg, Germany \\
\texttt{Martin.Braun@datev.de}, \texttt{Tanja.Doehler@datev.de}
\end{center}

\begin{abstract}
This study explores quantum and classical hybrid architectures for financial time‑series forecasting, focusing on Quantum Long Short‑Term Memory (QLSTM) networks and Quantum Reservoir Computing (QRC), using univariate and multivariate lag structures on real financial data. We assess how lag embeddings affect predictive accuracy and robustness. Data are encoded into quantum states via amplitude encoding, enabling efficient representation of normalized lagged observations under realistic qubit constraints. The recurrent dynamics of QLSTM and the reservoir of QRC are implemented as parameterized quantum circuits, while classical optimizers train the readout and, where applicable, variational circuit parameters. We benchmark quantum models against classical  LSTM and reservoir computing using common error like metrics. Our results show that, with suitable lag selection and amplitude encoding, quantum‑enhanced architectures match classical baselines in univariate settings and can modestly outperform them in multivariate regimes with correlated inputs, where expressive encodings are most beneficial. 
% We also analyze the impact of Q-Alchemy encoding on expressivity, noise sensitivity, and sample efficiency, and outline practical guidelines and open challenges for deploying PennyLane‑based quantum recurrent models on near‑term quantum hardware.

\end{abstract}


\section{Introduction}

Forecasting financial time series remains challenging due to non‑stationarity, heavy tails, regime shifts, and complex cross‑asset dependencies \cite{b1, b2}. Financial markets exhibit time‑varying volatility, structural breaks, and nonlinear interactions across instruments and sectors, which limit the effectiveness of purely linear models. At the same time, accurate forecasts of revenues, returns, and risk measures are central for pricing, risk management, and algorithmic trading across firms and institutions \cite{b3, b4, b5}.

The early twenty‑first century has seen rapid progress in quantum computing and quantum machine learning \cite{ab3,ab4,ab6}. Quantum algorithms exploit superposition and entanglement to realize high‑dimensional transformations that may be advantageous for certain computational tasks \cite{b6,b7}. Applying quantum mechanics to financial problems is not new: for example, the Black–Scholes–Merton framework can be mapped to a Schrödinger type equation, and entire financial markets can be modeled as quantum processes from which quantities such as covariance matrices arise naturally \cite{b8,b9,b10,c1,c2}. These links have motivated the exploration of quantum algorithms for option pricing, portfolio optimization, and risk analysis \cite{b12}.

Data driven methods based on machine learning, and deep learning in particular, have become central tools for financial forecasting \cite{b6, b11}. Recurrent neural networks (RNNs) are widely used to model temporal dependencies in asset returns, volatility, order book dynamics, and macroeconomic indicators \cite{c1}. Among these, Long Short‑Term Memory (LSTM) networks have emerged as a standard architecture because their gated memory cells mitigate vanishing gradients and enable the extraction of long range patterns from noisy, non stationary data \cite{ab2}. LSTMs, however, often require large parameter counts, extensive training, and substantial computational resources, especially in multivariate settings with many correlated series. In parallel, reservoir computing (RC) offers an alternative paradigm in which a fixed high‑dimensional recurrent “reservoir” generates rich nonlinear features, and only a simple readout is trained, providing favorable training cost and robustness properties \cite{ab1}.

Recent advances in quantum hardware and software have motivated quantum‑enhanced recurrent architectures that aim to leverage quantum state spaces as expressive feature maps for sequences. Quantum Long Short‑Term Memory (QLSTM) networks extend the LSTM paradigm by embedding parts of the recurrent computation into parameterized quantum circuits, potentially enabling more compact representations of complex temporal dependencies in a high‑dimensional Hilbert space \cite{ab2}. Quantum Reservoir Computing (QRC) generalizes classical RC by using a quantum system as the reservoir: classical inputs are encoded into quantum states, which evolve under fixed quantum dynamics; measurements at successive time steps provide features for a classical readout. The intrinsic complexity of quantum evolution can induce powerful nonlinear feature mappings that may better capture regime shifts and higher‑order dependencies in financial time series \cite{b12, b13}.

Prior work on quantum machine learning for time‑series has explored quantum recurrent models and quantum reservoirs for synthetic and small benchmark datasets, and quantum methods have been proposed for various financial tasks such as option pricing, portfolio optimization, and volatility estimation \cite{b5, ab6, b6, b7, b8}. Various researchers have worked with both classical and quantum algorithms to financial data, but many of these studies did not investigate comparable parameters between quantum and classical baselines, focus on classification tasks or signals instead of real financial time series, or analyze only a single quantum architecture in isolation.

In this study, we conduct a systematic, parameter matched comparison of QLSTM and QRC against LSTM and RC on financial time series, considering both univariate and multivariate lag structures in the analysis . We design quantum classical hybrid architectures for financial time‑series forecasting based on QLSTM and QRC, using amplitude encoding to embed lagged financial observations into quantum states under realistic qubit constraints, and we evaluate them in both univariate and multivariate settings.

This paper is organized as follows. Section II describes the financial dataset, lag construction, and preprocessing. Section III introduces the classical and quantum models, including QLSTM, QRC, and the amplitude encoding pipeline. Section IV presents the experimental setup and results for univariate and multivariate forecasting tasks. Section V concludes with a discussion of implications and directions for future work.

\section{Financial Data}\label{sec2}

\paragraph{Raw data.}
We investigate a financial time series forecasting task based on revenue-related measures for a larger set of products. For the present analysis, we retain 20 sufficiently complete and non-zero product series. Each retained product yields a univariate monthly time series of roughly 8 years (96 observations).

\paragraph{Synthetic data generation.}
Because the observed histories are short for studying long-range forecasting, we generate synthetic continuations that preserve key properties of the series. Many products exhibit a slowly varying mean with episodes of persistently high or low levels, so each product $d$ is modeled independently by a Gaussian process (GP) plus a two-state hidden Markov model (HMM) on the residuals.

For model fitting, the first 60 months are used as training data and the remaining 36 months are held out. Raw observations $y_{t,d}$ are transformed via
\begin{align}
\tilde{y}_{t,d} &= \max(y_{t,d},0), &
u_{t,d} &= \log(1+\tilde{y}_{t,d}), &
x_{t,d} &= \frac{u_{t,d}-\bar{u}_d}{s_d},
\end{align}
and time is rescaled to years as $\tau_t = (m_t-\min_t m_t)/12$.

The smooth component follows
\[
f_d(\tau) \sim \mathcal{GP}\!\left(\mu_d(\tau), k_d(\tau,\tau')\right),
\]
with a constant mean and an additive kernel combining trend, local/medium-scale variation, and seasonality (rational quadratic, Matérn, and periodic terms), implemented with GPyTorch \cite{b14, b15}. A two-state HMM captures level shifts on the residuals,
\[
x_t = f(\tau_t) + o_t + \varepsilon_t,\qquad
\varepsilon_t \sim \mathcal{N}(0,\sigma_\varepsilon^2),
\]
where $o_t$ is a state-dependent offset estimated using standard forward–backward and Baum–Welch algorithms~\cite{b16, b17}.

Synthetic continuations are obtained by evaluating the fitted GP on an extended time grid, sampling GP paths and HMM state sequences, adding Gaussian noise, and mapping back to the original scale via
\[
\hat{y}_{t,d} = \exp\!\left(s_d \hat{x}_{t,d} + \bar{u}_d\right)-1.
\]
This yields synthetic series that retain smooth trend, seasonality, and local level shifts.

\paragraph{Preprocessing and lag construction.}
For forecasting, we transform the sequential problem into a supervised learning task via lagged inputs. For a univariate process $\{x_t\}$, lagged feature vectors are of the form $(x_{t-1},\dots,x_{t-k})$ with target $y_t = x_t$ (or a future value $x_{t+h}$). This allows both classical and quantum models to capture temporal dependence and cyclical patterns based on the same lag structure as shown in table~\ref{lag} .

\begin{table}[h]
\centering
\caption{Lag based input output representation (univariate).\label{lag}}

\[
\left[
\begin{array}{c}
X\\
\hline
1\\
2\\
3\\
4\\
\vdots\\
n
\end{array}
\right]
\quad
\begin{array}{|c|c|c|c|c|}
\hline
X_{t-4} & X_{t-3} & X_{t-2} & X_{t-1} & Y_t \\
\hline
1 & 2 & 3 & 4 & 5 \\
\hline
2 & 3 & 4 & 5 & 6 \\
\hline
3 & 4 & 5 & 6 & 7 \\
\hline
4 & 5 & 6 & 7 & 8 \\
% \hline
% \vdots & \vdots & \vdots & \vdots & \vdots \\
\hline
\end{array}
\]

\end{table}
% In multivariate settings, lagged values of additional series (e.g., $y_{t-1}, z_{t-2}$)
Whereas, in multivariate we employ two columns of data \(X_{t0}\), e.g., for 2 columns of data \(X_{t1}\), the first column of data is embedded as \(X_{t0-3}, X_{t0-2}, X_{t0-1}, X_{t0-0}\) and \(X_{t0-4}\), and the second column of data is embedded as \(X_{t1-3}, X_{t1-2}, X_{t1-1}, X_{t1-0}\) and \(X_{t1-4}\), combined as a \(X_{t1-3}, X_{t1-2}, X_{t1-1}, X_{t1-0}, X_{t0-3}, X_{t0-2}, X_{t0-1}, X_{t0-0}\) and output is \( X_{t0-4}\) and \(X_{t1-4}\), of both columns are outputs, included to capture delayed cross-variable effects, yielding richer feature spaces. The lag length and structure must balance expressiveness and parsimony, too few lags may omit relevant information, whereas too many lead to high-dimensional, potentially collinear inputs and increased risk of overfitting. 

\begin{figure*}[htbp]
\centering
\subfloat[LSTM]{\includegraphics[width=0.48\linewidth, trim=20 20 20 20, clip]{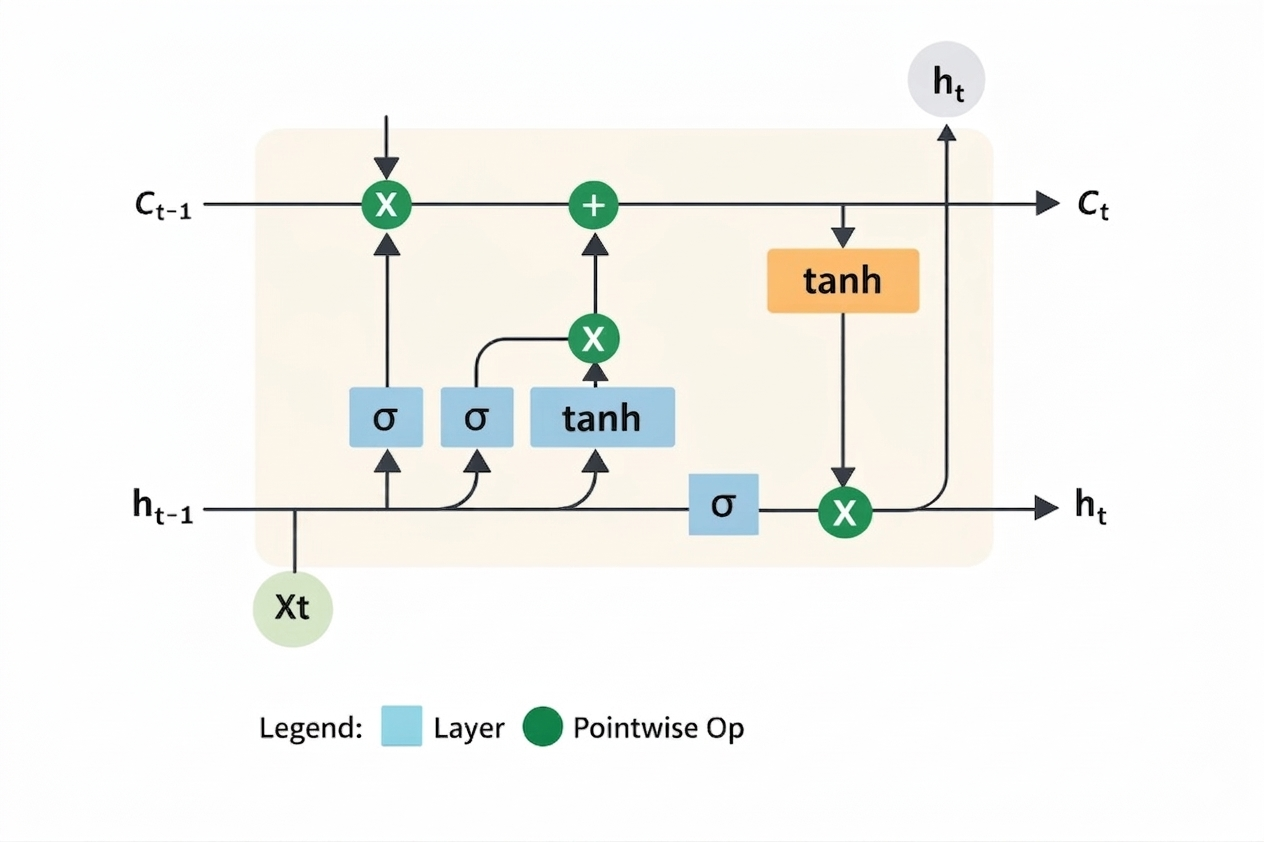}\label{lstm_a}}
\hfill
\subfloat[QLSTM VQCs]{\includegraphics[width=0.48\linewidth, trim=10 10 10 10, clip]{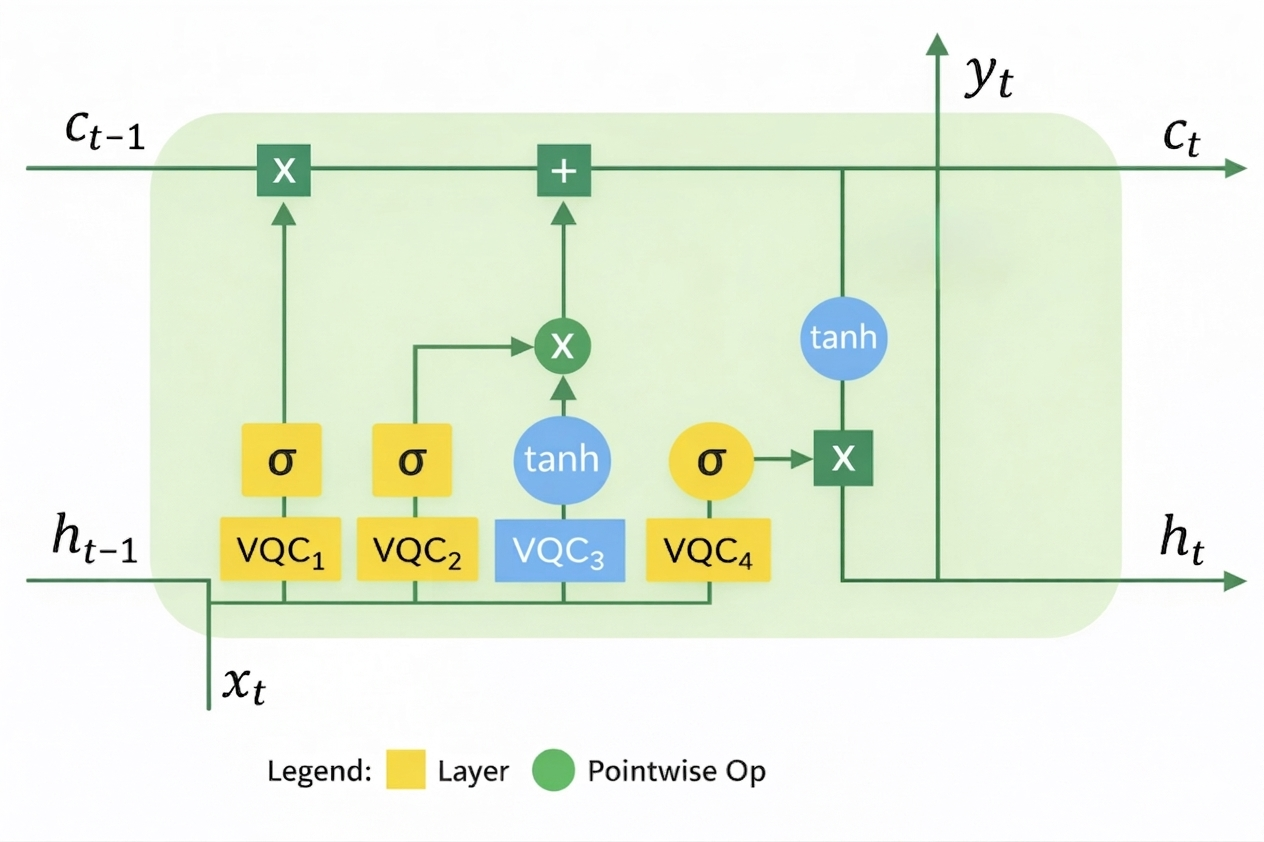}\label{qlstm_a}}
\caption{Comparison between classical LSTM and QLSTM with VQCs}
\label{fig:lstm_combined}
\end{figure*}

\paragraph{State preparation.}
To process classical data on quantum hardware, inputs must be encoded as quantum states \cite{b18}. In supervised learning, with data $D = \{(x_1,y_1),\dots,(x_n,y_n)\}$ and $y = f(x)$, each input $x_i$ is mapped to a quantum state $\ket{\psi_i}$, forming quantum data $((\ket{\psi_1},y_1),\dots,(\ket{\psi_n},y_n))$.

\paragraph{Amplitude encoding}
Beside pennyLane's amplitude encoding \cite{a18}, we employ Q‑Alchemy’s amplitude encoding as a data-loading strategy that embeds classical feature vectors directly into the probability amplitudes of an $n$-qubit state. Given a classical vector $\mathbf{x} \in \mathbb{R}^{2^n}$, Q‑Alchemy first normalizes it to $\tilde{\mathbf{x}} = \mathbf{x}/\|\mathbf{x}\|$ and prepares the target state
\[
\ket{\psi_{\mathbf{x}}} = \sum_{i=0}^{2^n-1} \tilde{x}_i \ket{i},
\quad
\sum_i |\tilde{x}_i|^2 = 1.
\]
This realizes exponential compression, as a $2^n$-dimensional feature vector is represented using $n$ qubits, making amplitude encoding an attractive front end for quantum feature maps \cite{b18, b19}.

Q‑Alchemy's amplitude encoding is implemented via a parameterized state-preparation circuit. Users specify an ansatz (e.g., layered single-qubit rotations and controlled rotations); parameters can either be chosen from a known state-preparation scheme (such as Möttönen-style preparation) or optimized with classical optimizers to maximize the fidelity between the prepared state and the target amplitudes. This integrates naturally into hybrid workflows: classical preprocessing normalizes and batches inputs, Q‑Alchemy compiles and executes the state-preparation circuit on the chosen backend, and the resulting quantum states feed into downstream quantum or hybrid models for regression or classification \cite{b19}.

\section{Methods }\label{sec3}
In this work, we consider a range of classical machine learning and quantum machine learning algorithms for time-series modeling. On the classical side, we employ Long Short-Term Memory networks (LSTM) and Reservoir computing (RC). On the quantum side, we consider Quantum Long Short-Term Memory networks (QLSTM) and Quantum Reservoir Computing (QRC).

\begin{algorithm}[H]
\caption{Quantum LSTM (QLSTM)}
\begin{algorithmic}[1]

\Require Input sequence $X = (x_1, x_2, ..., x_T)$
\Require Hidden size $H$, number of qubits $Q$, number of layers $L$
\Require Variational quantum circuits $\mathrm{VQC}_k(\cdot)$ for gates $k \in \{f,i,c,o\}$

\State Initialize hidden state $h_0 \leftarrow 0$
\State Initialize cell state $c_0 \leftarrow 0$

\For{$t = 1$ to $T$}

    \State Concatenate input and hidden state:
    \(
    v_t = [h_{t-1}, x_t]
    \)

    \State Classical encoding:
    \(e_t = W_{\mathrm{enc}} v_t \)
    \State \textbf{Quantum Gate Evaluation (shared structure)}

    \For{each gate $k \in \{f,i,c,o\}$}

        \State Encode $e_t$ into $Q$ qubits (e.g., amplitude encoding)

        \For{$l = 1$ to $L$}
            \For{each qubit $q$}
                \State Apply $H$
                \State Apply $R_Y(\theta^{(k)}_{l,q})$
            \EndFor
            \State Apply entanglement (CNOT ring)
        \EndFor

        \State Apply final rotations $Rot(\alpha, \beta, \gamma)$
        \State Measure:
        \(z^{(k)} = (\langle Z_1 \rangle, \dots, \langle Z_Q \rangle)
        \)

    \EndFor

\EndFor
\end{algorithmic}
\end{algorithm}

\begin{table}[t]
    \centering
    \caption{LSTM and QLSTM equations.}
    \label{eq1}
    \renewcommand{\arraystretch}{1.15}
    \begin{tabular}{>{\RaggedRight\arraybackslash}p{0.44\linewidth}|>{\RaggedRight\arraybackslash}p{0.44\linewidth}}
        \toprule
        \textbf{Classical LSTM} & \textbf{QLSTM} \\
        \midrule
        $f_t = \sigma(W_f v_t + b_f)$ & $f_t = \sigma(\mathrm{VQC}_1(v_t))$ \\
        $i_t = \sigma(W_i v_t + b_i)$ & $i_t = \sigma(\mathrm{VQC}_2(v_t))$ \\
        $\tilde{C}_t = \tanh(W_C v_t + b_C)$ & $\tilde{C}_t = \tanh(\mathrm{VQC}_3(v_t))$ \\
        $c_t = f_t \cdot c_{t-1} + i_t \cdot \tilde{C}_t$ & $c_t = f_t \cdot c_{t-1} + i_t \cdot \tilde{C}_t$ \\
        $o_t = \sigma(W_o v_t + b_o)$ & $o_t = \sigma(\mathrm{VQC}_4(v_t))$ \\
        $h_t = o_t \cdot \tanh(c_t)$ & $h_t = o_t \cdot \tanh(c_t)$ \\
        \bottomrule
    \end{tabular}
\end{table}

\textbf{Long Short-Term Memory}
(LSTM) networks are a particular type of RNN, specifically designed to handle long-range dependencies in sequential information. They were developed to solve the problem of vanishing gradients, which affects regular RNNs as illusrated in \ref{lstm_a}. The LSTM cell at time t consists of a memory cell $c_t$ and a hidden state $h_t$, as well as gates that control the flow of information: an input gate $i_t$, a forget gate $f_t$, and an output gate $\tilde{C}_t$ as shown in the LSTM side of eq.~\ref{eq1}. Whereas, $w$ weights at respective gates neurons, $v_t$ input at current timestamp and $b$ biases. Training is performed via backpropagation through time, allowing end-to-end optimization of both short and long term dependencies for tasks such as sequence prediction, language modeling, and time series forecasting \cite{ab2}.

\begin{figure}[htbp]
    \centering
    \subfloat[Variational Layer]{\includegraphics[width=0.60\linewidth]{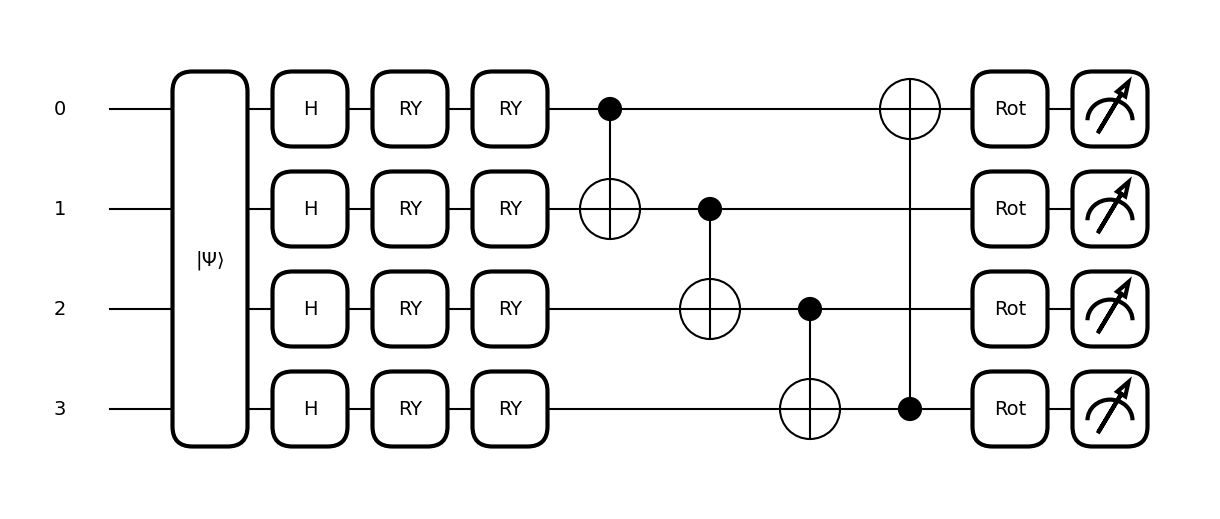}\label{fig:vqc_sub_a}}
    \hfill
    \subfloat[QLSTM VQC]{\includegraphics[width=0.35\linewidth]{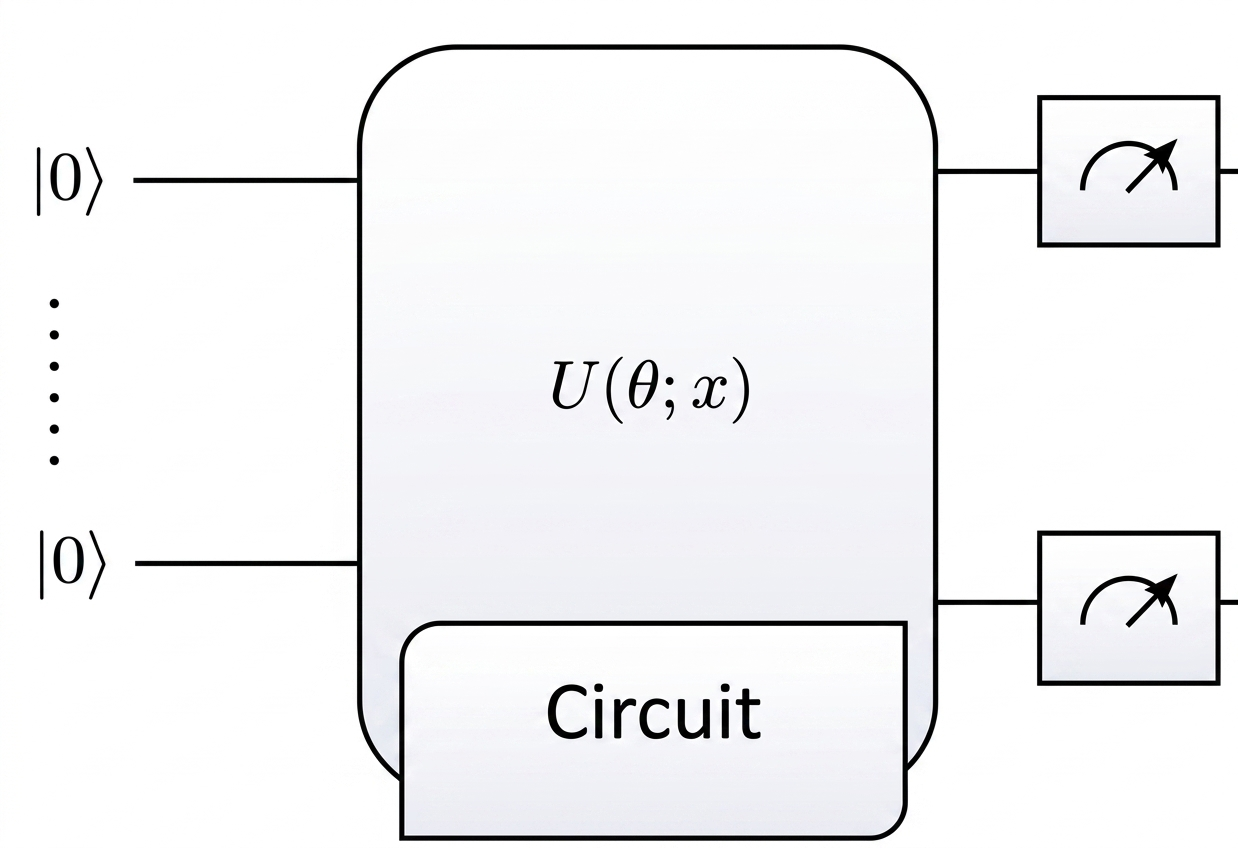}\label{fig:vqc_sub_b}}
    \caption{Variational qauntum classifer (a) Variational Layer and (b) Variational qauntum classifer schemtic}
    \label{fig:vqc_combined}
\end{figure}

\textbf{Quantum Long Short-Term Memory}
(QLSTM) merges the temporal modeling capabilities of classical LSTMs with the representational advantages of quantum computation by implementing LSTM-like gating and memory mechanisms within parametrized quantum circuits as shown in eq.~\ref{eq1}. We employ the Q-Alchemy amplitude encoding to encode sequential data into quantum states, process them with a variational quantum classifier (VQC) that realizes analogues of input, forget, and output gates, and then interface them with classical components for loss evaluation and parameter updates in Fig \ref{qlstm_a} \& \ref{fig:vqc_sub_b}. These architectures aim to exploit quantum parallelism to more expressively handle long-range dependencies in sequences while maintaining the stability properties characteristic of LSTMs \cite{ab2, b18}.

\textbf{Variational quantum circuits}
(VQCs), also called parameterized quantum circuits, are quantum gate circuits with tunable parameters \(U(x)\) and form a core component of many quantum computing models. Computation proceeds by applying a sequence of quantum gates that transform the states of qubits the basic units of quantum information and then measuring the outputs at the end of the circuit as shown in Fig.~\ref{fig:vqc_sub_b} \cite{b18}. 
To encode classical data into a quantum state, we employ Q-Alchemy amplitude encoding and a variational layer. For the variational layer, we use a combination of $\mathrm{CNOT}$ and rotation gates; the qubits are entangled and rotated toward the target state. The VQC trainable parameters are the three rotation angles $\alpha$, $\beta$, and $\gamma$ associated with the $R_X$, $R_Y$, and $R_Z$ gates. The expectation value of each qubit is measured with respect to Pauli Z ($\sigma_z$) to transform quantum information into classical information as shown in Figs.~\ref{fig:vqc_sub_a} \& \ref{fig:vqc_sub_b}. The model is optimized classically, and these learnable parameters are updated iteratively via gradient descent through $V(\theta)$ refer eq.~\ref{eq2}\cite{b18}.

\textbf{\begin{equation}
\left| \psi(x: \theta)\right\rangle = U(\theta) \left| \phi(x)\right\rangle
\label{eq2}
\end{equation}}

\begin{figure*}[!t]
\centering
\subfloat[RC]{\includegraphics[width=0.40\linewidth, trim=20 20 20 20, clip]{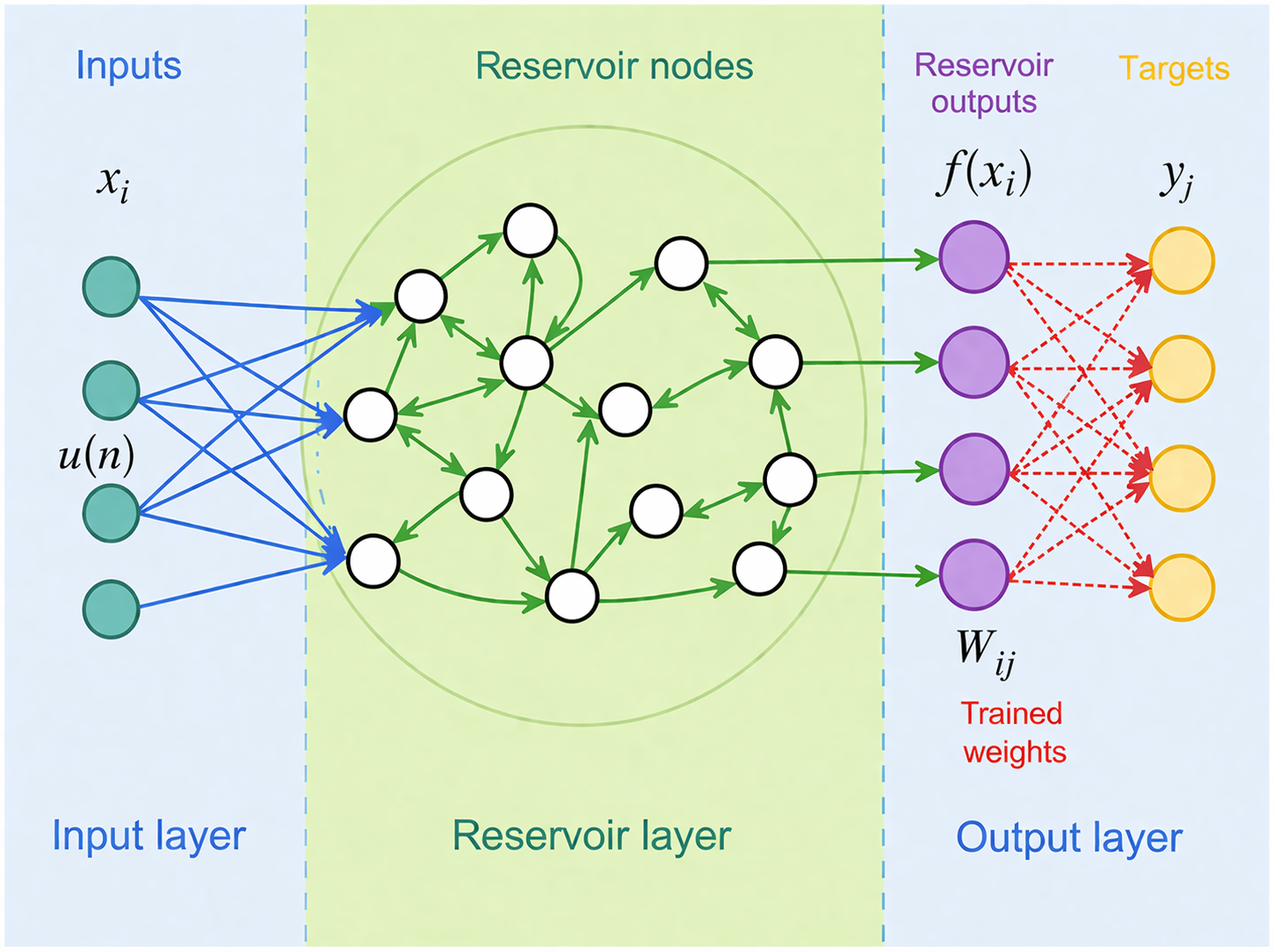}\label{rc}}
\hfill
\subfloat[QRC]{\includegraphics[width=0.40\linewidth, trim=10 10 10 10, clip]{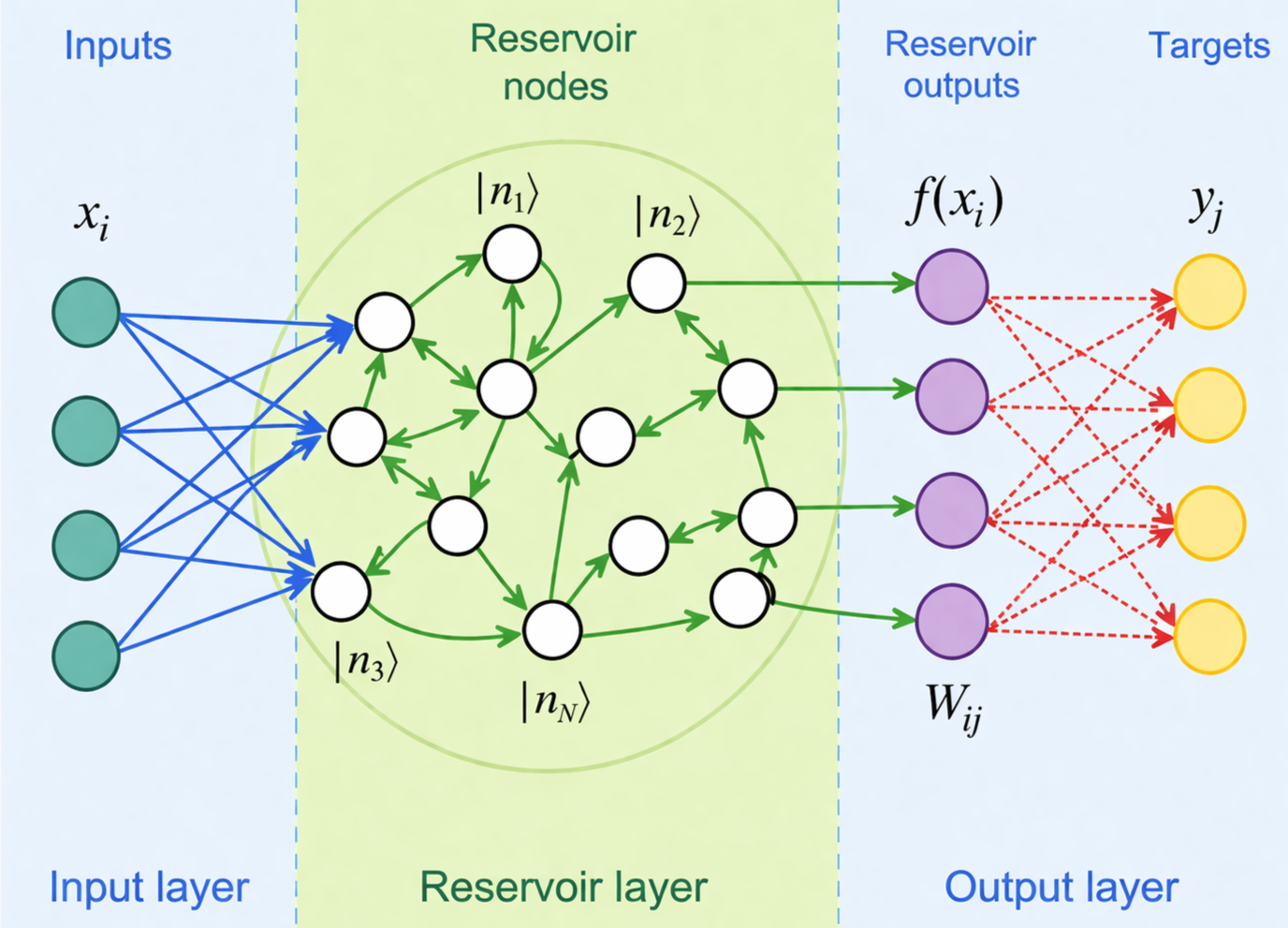}\label{qrc}}
\caption{Comparison between classical RC and QRC architectures.}
\label{fig:qrc_combined}
\end{figure*}

\textbf{Reservoir Computing}
(RC) is a type of recurrent neural network where only the linear readout is adapted during training, while the recurrent reservoir is left unmodified after random initialization. This type of reservoir is often a high-dimensional, sparsely connected dynamical system with a nonlinear mapping of the input sequences into a large state space. This approach avoids many of the optimization problems of traditional recurrent neural networks, as there is no gradient flow through the internal weights during training. This allows for fast, convex optimization of the readout using traditional linear regression. Therefore, reservoir computing is particularly useful for tasks in time series modeling, system identification, and signal processing, where speed of training and computational cost are important. Functionally, the reservoir can be viewed as a temporal kernel that implicitly includes nonlinear mappings as well as memory of past inputs. Therefore, by properly adapting the global parameters of the system, the reservoir can be designed to possess certain desirable properties, such as fading memory and echo state, which are critical for stable sequence processing. This approach is surprisingly effective, considering the relative simplicity of the approach, as it has been demonstrated for a wide range of tasks, from chaotic time series prediction, speech recognition, and control problems. Furthermore, there is flexibility in the physical implementation of the reservoir, as it can be implemented both as a software network and as a hardware network, which is promising for energy-efficient sequence processing as illustrated in Fig.~\ref{rc} \cite{b8}.

% \begin{algorithm}[H]
% \caption{Training Quantum LSTM}
% \begin{algorithmic}[1]

% \Require Training data $(X_{train}, y_{train})$
% \Require Test data $(X_{test}, y_{test})$
% \Require Learning rate $\eta$, epochs $E$

% \State Initialize model parameters
% \State Initialize Adam optimizer
% \State Define loss $\mathcal{L} = \text{MSE}$

% \For{$epoch = 1$ to $E$}

%     \State $\hat{y} \leftarrow \text{Model}(X_{train})$
%     \State Compute loss: \(
%     \mathcal{L} = \frac{1}{N}\sum_i (\hat{y}_i - y_i)^2
%     \)
%     \State Backpropagate gradients
%     \State Update parameters via Adam

% \EndFor

% \State Evaluate on test set
% \(\text{MSE}, \text{MAE}
% \)

% \end{algorithmic}
% \end{algorithm}

\textbf{Quantum reservoir computing}
% Quantum reservoir computing (QRC) adapts the reservoir computing paradigm to quantum systems by usin
(QRC) uses a fixed, randomly initialized quantum dynamical system as a reservoir and trains only a classical readout layer. Inputs are encoded into quantum states, evolved by an appropriate Hamiltonian or circuit, and measured to yield high‑dimensional, nonlinear features. A fixed nonlinear quantum “reservoir” provides memory and nonlinear feature extraction, while the linear readout is trained with regularized regression as shown in Fig.~\ref{qrc}. Correct amplitude encoding is essential: state preparation must yield a normalized vector of length \(2^n\) for \(n\) qubits; naïve zero padding biases results and wastes Hilbert space capacity. We replace padding with a fixed random projection to a higher dimensional feature space, followed by a single normalization before state preparation, consistent with established QRC literature and PennyLane amplitude encoding guidance \cite{b8, b13}.

Given input features \(x^{(t)} \in \mathbb{R}^{d_x}\), a bias vector \(b \in \mathbb{R}^{n}\), and the previous measurement state \(r^{(t-1)} \in \mathbb{R}^{3n}\), we first build a per-qubit input, and compress the reservoir memory via the following equations \cite{Jaeger2001}:

\begin{align}
x^{(t)}_i &= x^{(t)}_{\,i \bmod d_x} + b_i, \quad i=1,\dots,n,\\
m^{(t-1)}_i &= w_X\,\langle X_i\rangle^{(t-1)} 
            + w_Y\,\langle Y_i\rangle^{(t-1)} \notag\\
            &\quad + w_Z\,\langle Z_i\rangle^{(t-1)}, 
            \quad i=1,\dots,n,\\
v^{(t)}_i &= \lambda\,x^{(t)}_i + (1-\lambda)\,m^{(t-1)}_i, 
            \quad \lambda \in [0,1].
\end{align}

Here, $\langle X_i\rangle$, $\langle Y_i\rangle$ and $\langle Z_i\rangle$ are the measurement results of the last iteration step.
The lifted fearture vector is multiplied with a random projection \(P\) and is then normalized.
In the quantum circuit, for each layer \(l=1,\dots,L\), we apply independent single-qubit rotations and weak ring entanglement:

\begin{equation}
\begin{aligned}
U^{(l)} \;=\;& \Bigg(\prod_{i=1}^{n} R(\theta_{l,i},\phi_{l,i},\lambda_{l,i})_i \Bigg) \\
&\cdot \Bigg(\prod_{i=1}^{n-1} \mathrm{CNOT}_{i\rightarrow i+1}\Bigg)
\cdot \mathrm{CNOT}_{n\rightarrow 1}
\end{aligned}
\end{equation}

where \(R(\theta,\phi,\lambda)\) denotes a fixed, random single-qubit rotation. The overall operator \(U = U^{(L)}\cdots U^{(1)}\) yields \(\lvert \psi^{(t)} \rangle = U\,\lvert \psi^{(t)}_0 \rangle\).
We measure, for each qubit, the expectation values of three observables and stack them:

\begin{equation}
\begin{aligned}
r^{(t)} = \big[\, 
&\langle X_1\rangle,\dots,\langle X_n\rangle, \\
&\langle Y_1\rangle,\dots,\langle Y_n\rangle, \\
&\langle Z_1\rangle,\dots,\langle Z_n\rangle 
\,\big]^\top \in \mathbb{R}^{3n}.
\end{aligned}
\end{equation}

Predictions are obtained via a linear readout:

\begin{equation}
\hat{y}^{(t)} \;=\; W_{\mathrm{out}}^\top\,r^{(t)}.
\end{equation}
The reservoir (rotations, entanglement, projection, and bias) remains fixed after initialization; only \(W_{\mathrm{out}}\) is trained classically. With \(R\in\mathbb{R}^{T\times 3n}\) collecting all time-step measurements and \(Y\in\mathbb{R}^{T\times 1}\) the targets, we solve ridge regression:
\begin{equation}
W_{\mathrm{out}} \;=\; \big(R^\top R + \lambda_{\mathrm{reg}} I\big)^{-1} R^\top Y,
\end{equation}
with \(\lambda_{\mathrm{reg}} > 0\) controlling regularization. Dimensional consistency is enforced (\(T\) time steps in \(R\) and \(Y\) must match).

\begin{algorithm}[H]
\caption{Quantum Reservoir Computing with Amplitude Encoding}
\begin{algorithmic}[1]

\Require Input sequence $x_{1:T}$, number of qubits $n$, layers $L$, leakage rate $\alpha$
\Require Random projection matrix $P \in \mathbb{R}^{2^n \times d}$
\State Initialize reservoir parameters $\theta$
\State Initialize previous state $s_0 = 0 \in \mathbb{R}^{3n}$

\For{$t = 1$ to $T$}

    \State \textbf{Memory Compression} $m_{t-1} \gets f_{\text{compress}}(s_{t-1})$

    \State \textbf{Input Mapping} $\tilde{x}_t[i] \gets x_t[i \bmod d_x] + b_i$

    \State \textbf{Leaky Integration} $v_t \gets \alpha \tilde{x}_t + (1-\alpha)m_{t-1}$

    \State \textbf{Feature Lifting} $u_t \gets f_{\text{lift}}(v_t)$

    \State \textbf{Projection to Hilbert Space} $a_t \gets P u_t$

    \State \textbf{Normalization}
    \State $\psi_t \gets a_t / \|a_t\|$

    \State \textbf{Quantum State Preparation} $|\psi_t\rangle = \sum_i \psi_{t,i} |i\rangle$

    \State \textbf{Reservoir Evolution}
    \For{$l = 1$ to $L$}
        \For{$i = 1$ to $n$}
            \State Apply $Rot(\theta_{l,i},\phi_{l,i},\lambda_{l,i})$
        \EndFor
        \State Apply ring entanglement using CNOT gates
    \EndFor

    \State \textbf{Measurement} $s_t \gets [\langle X_i \rangle, \langle Y_i \rangle, \langle Z_i \rangle]_{i=1}^n$

\EndFor

\State \textbf{Readout Training} $W_{out}$ using regression on collected states $s_t$

\end{algorithmic}
\end{algorithm}

\section{Results \& Discussions}\label{sec4}

In this study, we use both CPU‑ and QPU‑based algorithms for time‑series prediction. The training models are implemented in Python 3 using PyTorch and the scikit‑learn library. We access Q‑Alchemy via its API to encode classical data into quantum states and integrate this encoding into the PennyLane framework. In this work, we splited the data into 80-20 for training and testing. 

We compare the performance and capabilities of QLSTM and QRC with their classical counterparts across univariate and multivariate time‑series prediction tasks.

% We use 2 configuations of the data: univariate and multivariate. In the univariate case, the input data is encoded as \(X_{t-3}, X_{t-2}, X_{t-1}, X_{t-0}\) and \(X_{t-4}\) is the output. Using amplitude encoding, we embed into 2 qubits. Whereas, in multivariate we employ more than one column of data \(X_{t0}\), e.g., for 2 c{t1-1}, X_{t1-0}\) and \(X_{t1-4}\), combined as a \(X_{t1-3}, X_{t1-2}, X_{t1-1}, X_{t1-0}, X_{t0-3}, X_{t0-2}, X_{t0-1}, X_{t0-0}\) anolumns of data \(X_{t1}\), the first column of data is embedded as \(X_{t0-3}, X_{t0-2}, X_{t0-1}, X_{t0-0}\) and \(X_{t0-4}\), and the second column of data is embedded as \(X_{t1-3}, X_{t1-2}, X_d output is \( X_{t0-4}\) and \(X_{t1-4}\), of both columns are outputs. 

\begin{figure*}[htbp]
    \centering
    \subfloat[LSTM (Univariate)]{\includegraphics[width=0.4\linewidth]{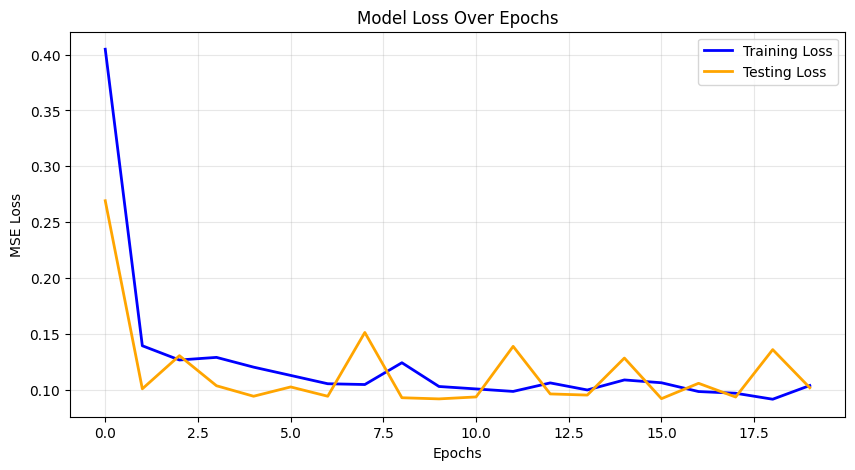}\label{uni-lstm}}\hfill
    \subfloat[QLSTM (Univariate)]{\includegraphics[width=0.4\linewidth]{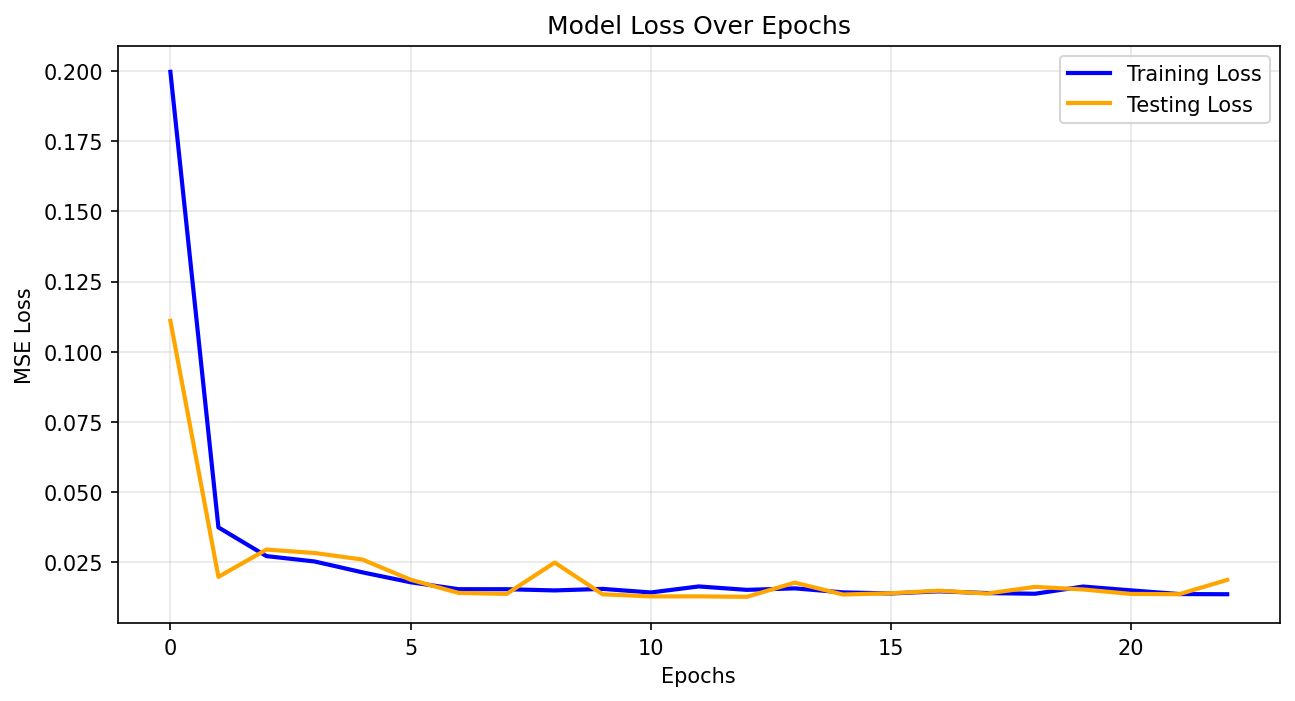}\label{uni-qlstm}}

    \subfloat[LSTM (Multivariate)]{\includegraphics[width=0.4\linewidth]{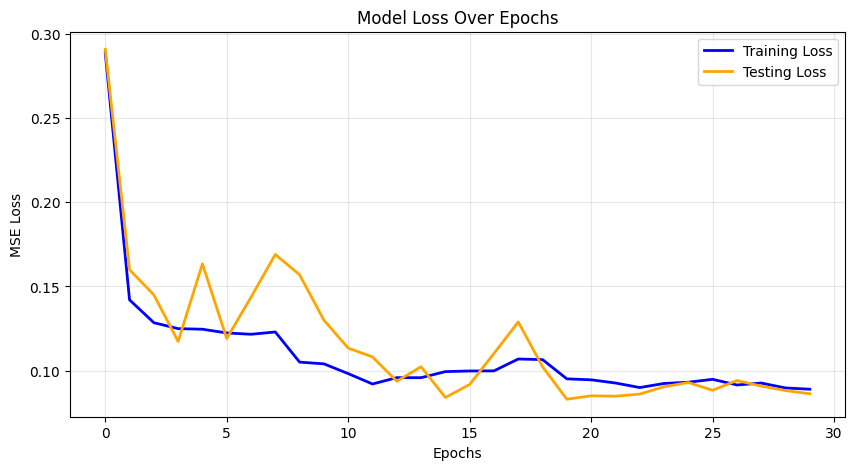}\label{multi-lstm}}\hfill
    \subfloat[QLSTM (Multivariate)]{\includegraphics[width=0.4\linewidth]{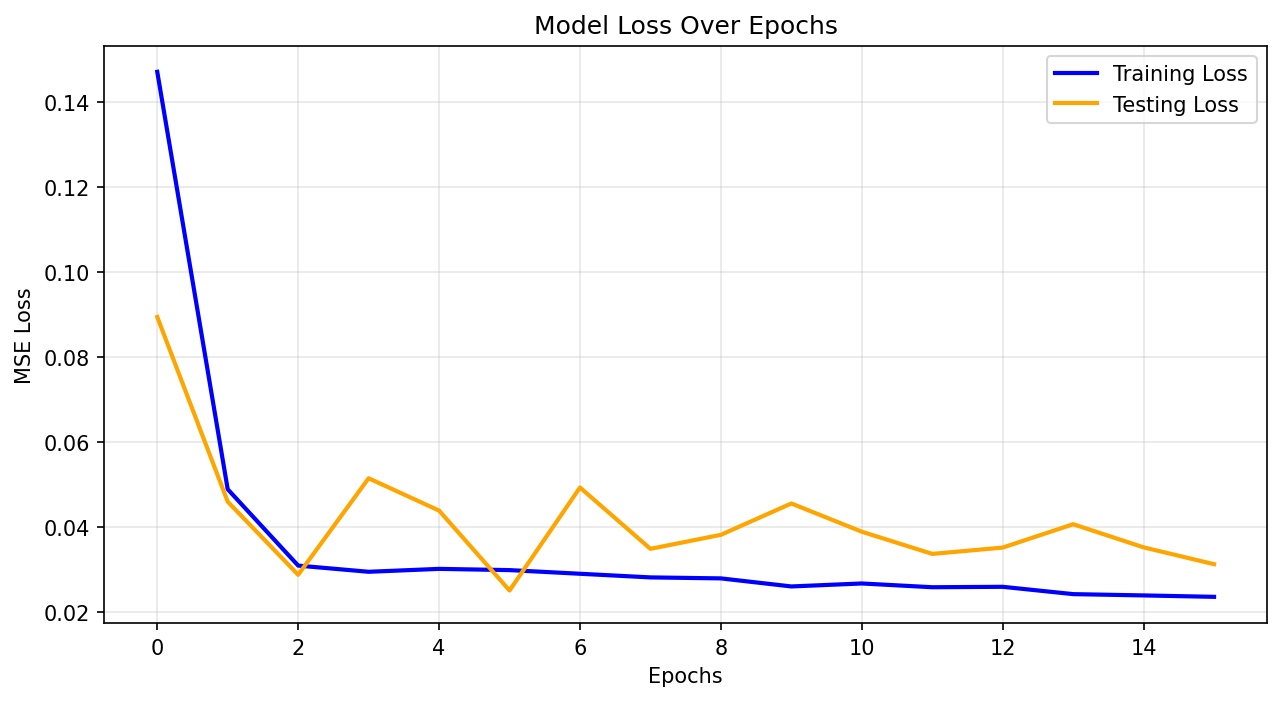}\label{multi-qlstm}}

    \caption{Loss comparison of LSTM and QLSTM models for univariate and multivariate settings.}
    \label{fig:loss_2x2}
\end{figure*}

\begin{figure*}[htbp]
    \centering
    \subfloat[LSTM (Univariate)]{\includegraphics[width=0.47\linewidth]{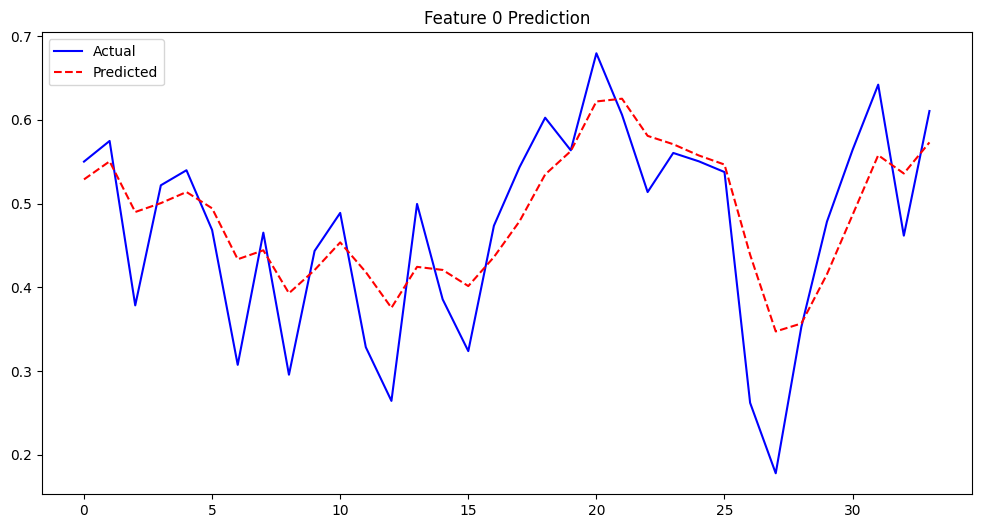}\label{uni-lstm-p}}\hfill
    \subfloat[QLSTM (Univariate)]{\includegraphics[width=0.48\linewidth]{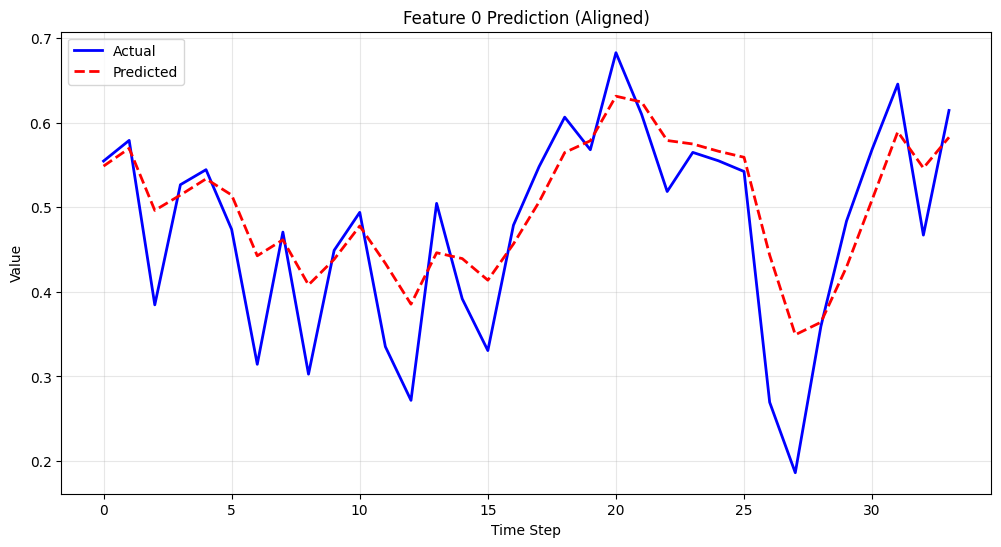}\label{uni-qlstm-p}}

    \subfloat[LSTM (Multivariate)]{\includegraphics[width=0.52\linewidth]{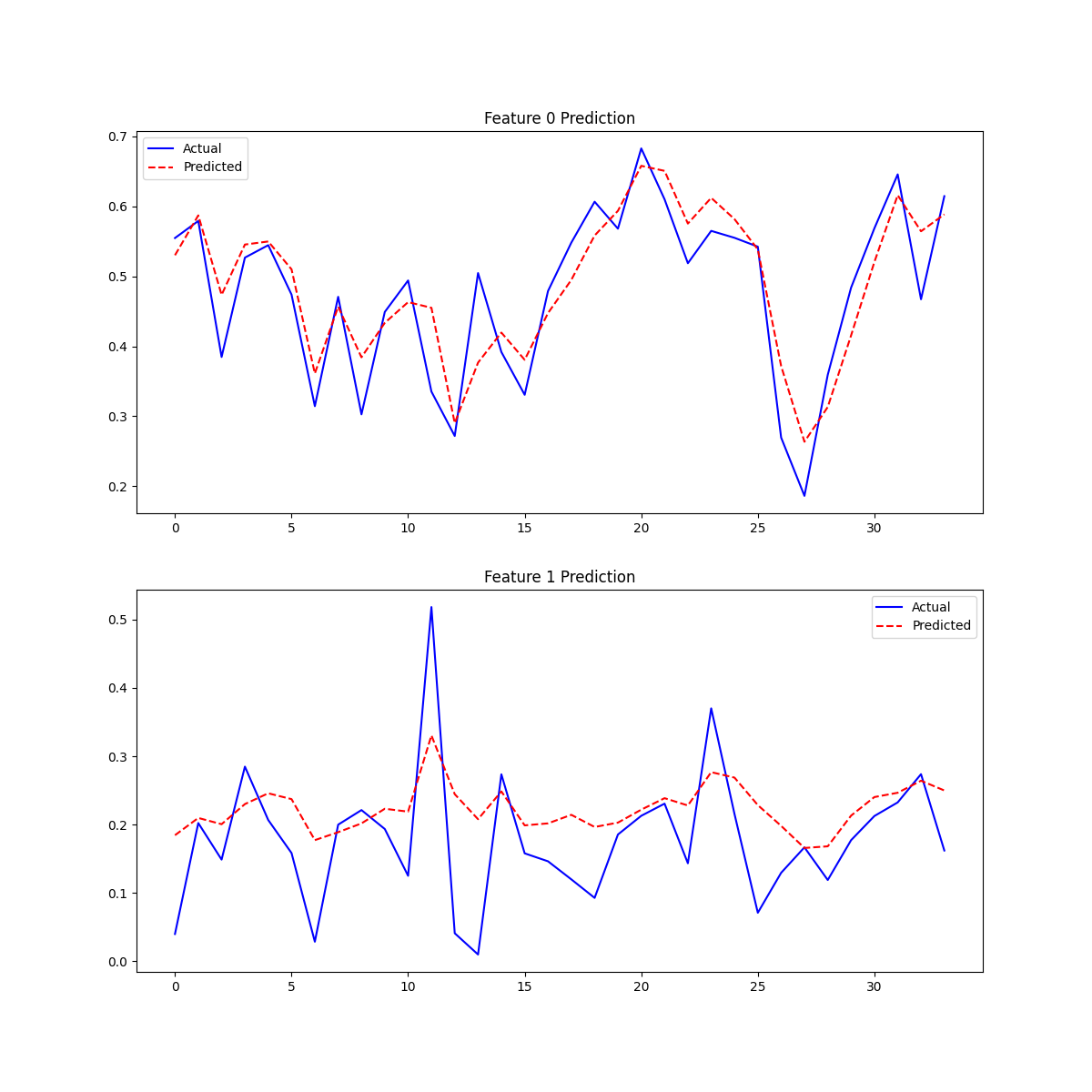}\label{multi-lstm-p}}\hfill
    \subfloat[QLSTM (Multivariate)]{\includegraphics[width=0.48\linewidth]{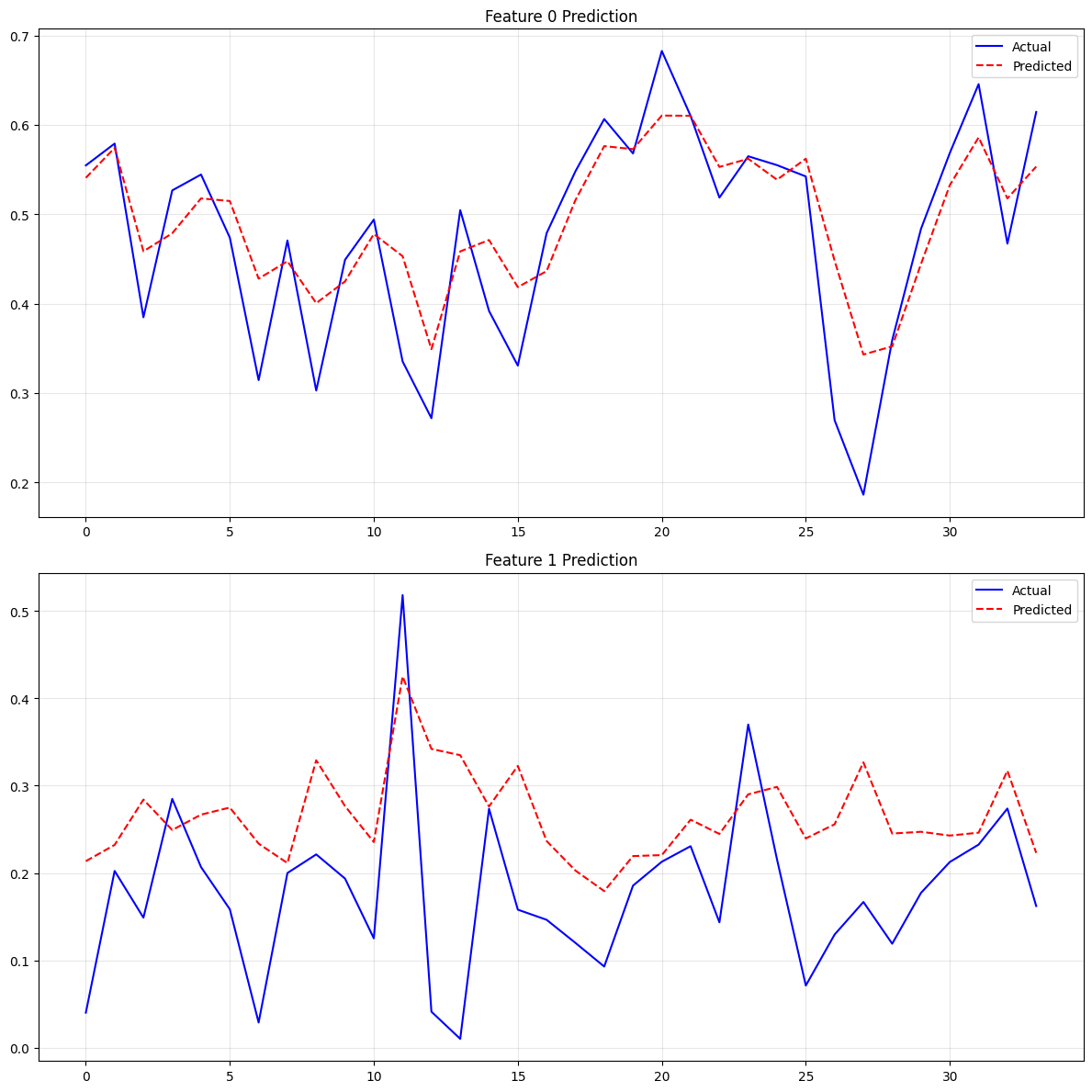}\label{multi-qlstm-p}}

    \caption{Prediction comparison of LSTM and QLSTM models for univariate and multivariate settings.}
    \label{fig:pred_2x2}
\end{figure*}

For our QLSTM model, we use four VQCs to amplitude-encode the qubits and a variational layer with depth 2. For the QLSTM and LSTM we kept the same hidden states and learning rate. This ensures a fair comparison of performance between the LSTM and QLSTM models.
In all experiments, we explicitly measure the performance and capacity of the model. The QLSTM employs VQC circuit to amplitude‑encode the qubits, followed by a variational layer with depth 2, resulting in a total of trainable parameters. The classical baseline is an LSTM which closely matches the QLSTM in terms of parameter count. This alignment ensures that performance differences primarily reflect representational differences between quantum and classical models rather than differences in model size.

In the univariate configuration, where the objective is to forecast a single time series from its past observations, the two models perform similarly overall. The QLSTM attains slightly lower RMSE values than the LSTM, indicating a small but consistent improvement in pointwise prediction accuracy. This suggests that the quantum recurrent cell introduces additional expressive capacity, although this advantage remains modest in the single channel time series setup.

The learning curves in Figs.~\ref{uni-lstm} \& \ref{uni-qlstm} illustrate this behavior. Both the LSTM and QLSTM exhibit rapid initial loss reduction, followed by convergence to a stable plateau without pronounced indications of underfitting or overfitting. The QLSTM converges to marginally lower final loss values than the LSTM; however, the gap between the two remains narrow. From a practical standpoint, this difference may not always justify the overhead associated with integrating quantum components into an otherwise classical modeling pipeline.

A qualitative comparison in Figs.~\ref{uni-lstm-p} \& \ref{uni-qlstm-p} confirms these observations. In both cases, the predicted univariate trajectories closely follow the ground truth: dominant structures such as peaks, troughs, and short‑term fluctuations are captured reliably by both architectures. In some segments, QLSTM predictions appear slightly better aligned with the true signal, for example through smoother transitions or reduced phase offsets, but these effects are subtle. Overall, the univariate results indicate that, under current device noise and resource limitations, the use of quantum recurrent layers is not crucial when dealing with single channel time series.

The multivariate experiments, in which multiple correlated time series are modeled jointly, reveal a more substantial difference between the two approaches. In this setting, the QLSTM consistently outperforms the LSTM across all considered error metrics. The reduction in prediction error is observed on both the training and test sets, suggesting that the performance gain generalizes beyond the training data and is not merely due to overfitting see Figs.~\ref{multi-lstm} \& \ref{multi-qlstm}.

The training dynamics highlight this advantage. While both models converge stably, the QLSTM reaches a noticeably lower loss plateau than the LSTM, yielding a clearly visible separation of the loss curves by the end of training see Figs.~\ref{multi-lstm}\& \ref{multi-qlstm}. This behavior indicates that the QLSTM is better suited to exploit the increased dimensionality of the input and the richer correlation structure across channels. In particular, the quantum recurrent mechanism appears more capable of encoding nonlinear interdependencies that emerge only when several variables are processed simultaneously.

The qualitative analysis in Figs.~\ref{multi-lstm-p} \& \ref{multi-qlstm-p} further supports this conclusion. Across all channels, the QLSTM predictions exhibit closer agreement with the ground truth, preserving both the amplitude and phase of the temporal patterns. Cross channel structures such as lagged interactions or coupled oscillations between variables are reproduced more faithfully by the QLSTM. By contrast, the LSTM occasionally exhibits damped dynamics or small phase shifts in some channels. Taken together, these qualitative and quantitative results indicate that the QLSTM derives a pronounced benefit from multivariate inputs with significant multi channel correlations.

In the proposed QRC framework, the reservoir is realized as the evolution of input‑dependent quantum states within a fixed quantum dynamical system. At each time step, the classical input is encoded into the quantum system, which then undergoes unitary evolution and generates a high‑dimensional quantum state. A linear readout (LR QRC) is trained on measurement outcomes of this state to produce intermediate predictions. These intermediate outputs are subsequently passed to a neural‑network‑based QRC readout (NN QRC), which refines the representation and yields the final, measurement‑based forecasts. In this way, the QRC architecture combines the complex dynamics of a quantum reservoir with both linear and nonlinear classical readout stages.

To ensure a fair comparison, the effective reservoir size and the readout complexity are kept comparable between the classical Reservoir Computing (RC) and QRC models. In the classical RC model, the reservoir consists of \(R_C\) units with fixed random connectivity, and only the readout weights are trained. First, a linear regression (LR RC) readout is fitted to the reservoir states, and its outputs are used to generate intermediate predictions. These intermediate outputs are then provided as input to an RC‑based neural network (NN RC), which produces the final forecasts. In the QRC model, the reservoir is given by the evolution of input‑dependent quantum states within the fixed quantum dynamical system, and the output of the LR QRC serves as the input to the NN QRC, which then produces the final, measurement‑based predictions.

\begin{figure*}
    \centering
    \subfloat[Univariate QRC]{\includegraphics[width=0.45\linewidth]{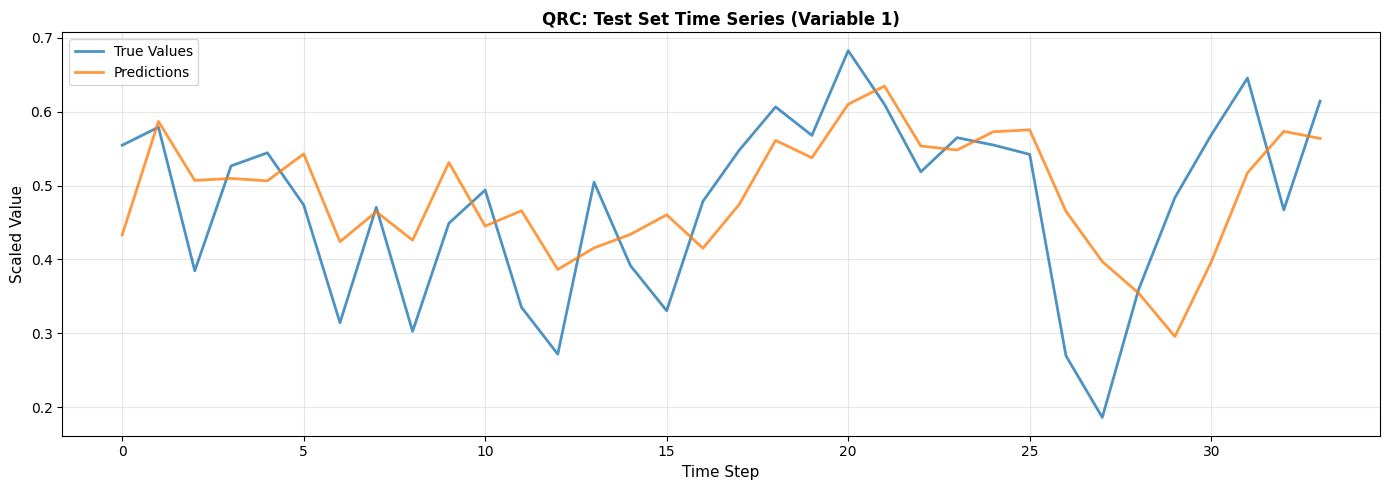}\label{fig:qrc_uni}}\hfill
    \subfloat[Univariate NN QRC]{\includegraphics[width=0.45\linewidth]{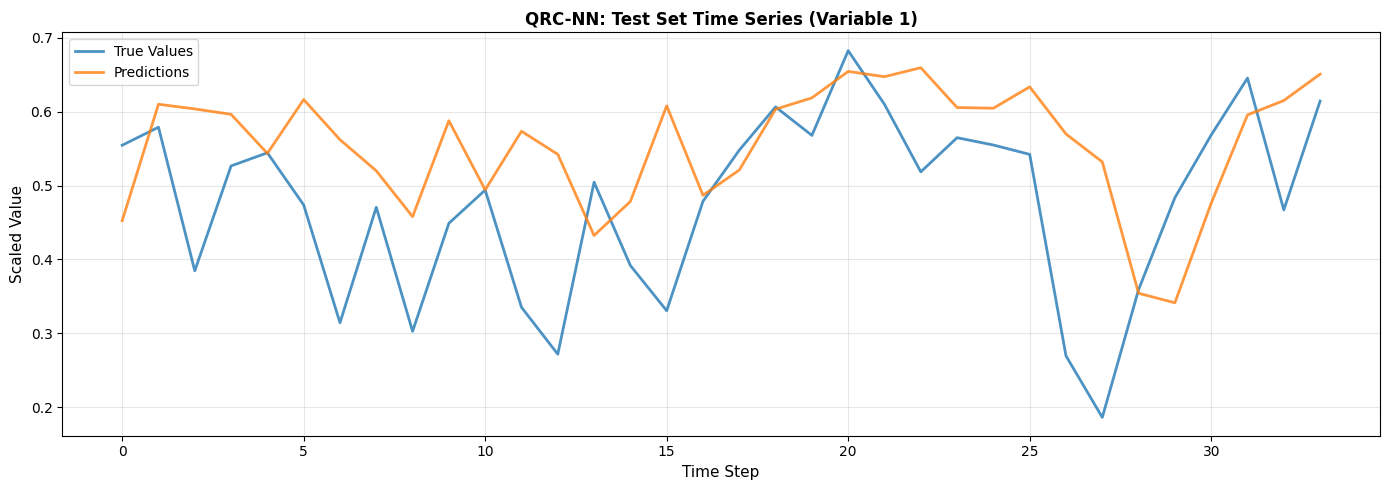}\label{fig:qrc_uni_linear}}

    \subfloat[Univariate RC]{\includegraphics[width=0.45\linewidth]{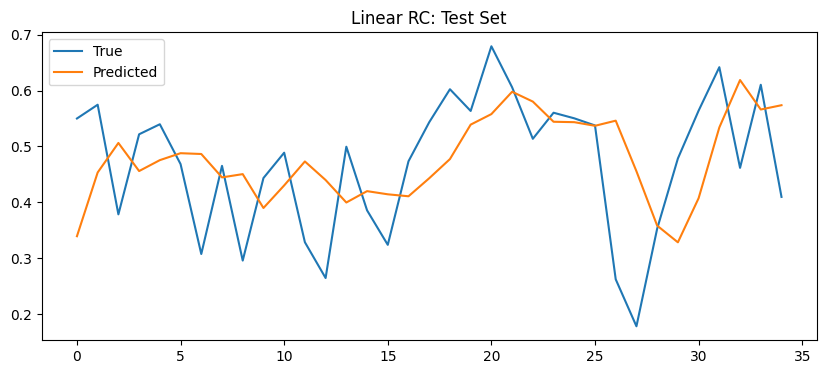}\label{fig:rc_uni_linear}}\hfill
    \subfloat[Univariate NN RC]{\includegraphics[width=0.45\linewidth]{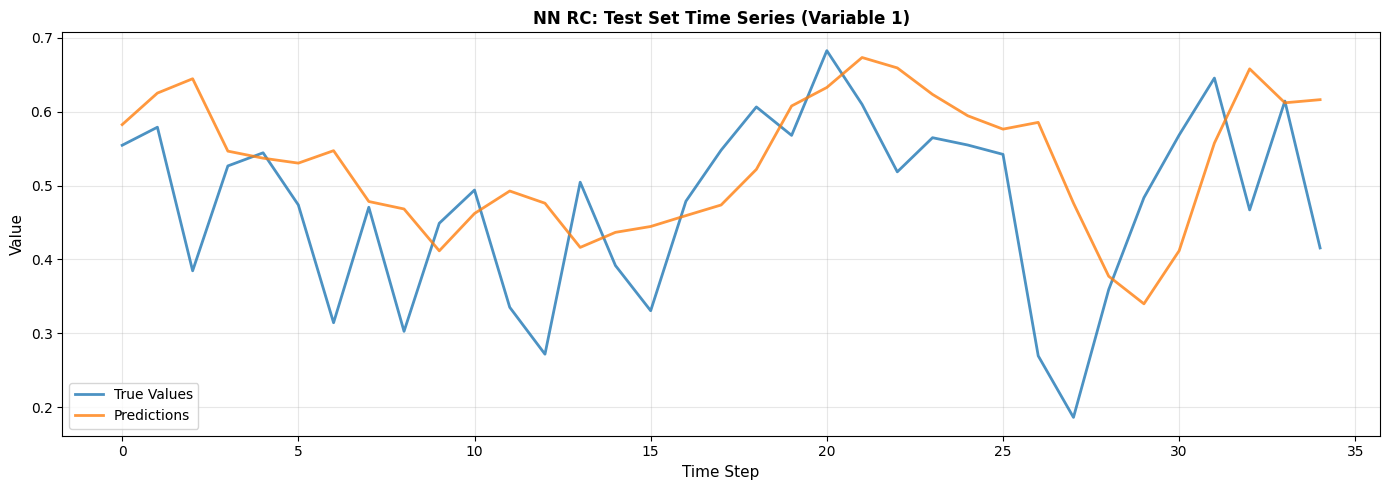}\label{fig:nn_rc_uni}}

    \caption{QRC and RC comparison on univariate data.}
    \label{fig:qrc_2x2_alt}
\end{figure*}

In the univariate case, the QRC variants show slightly higher RMSE values than the corresponding classical RC models, and only marginal gains in pseudo‑accuracy see Fig.~\ref{fig:qrc_uni}, \ref {fig:qrc_uni_linear} ,\ref{fig:rc_uni_linear} \& \ref{fig:nn_rc_uni}. Overall, the performance gap between RC and QRC in this setting is small. This indicates that, for time series with relatively simple temporal structure, the nonlinear mixing achieved by a classical reservoir already provides a sufficiently expressive high‑dimensional feature space Fig. \ref{fig:qrc_2x2_alt}. In such cases, the additional representational power of the quantum reservoir does not lead to a marked improvement in forecasting quality.

Visual inspection of the univariate trajectories supports this interpretation. As illustrated in Fig.~\ref{fig:qrc_2x2_alt}, both RC and QRC (including LR QRC and NN QRC) generate predictions that closely track the ground truth, successfully capturing the dominant patterns such as peaks and troughs. The predicted curves from RC and QRC nearly coincide, and the magnitude of the residuals between actual and predicted values is comparable across all variants. These observations suggest that, in the univariate regime, classical reservoir computing already captures the essential dynamics, while the quantum extension yields at most modest benefits under current hardware constraints

\begin{figure*}[!t]
    \centering
    \subfloat[Univariate losses and accuracies]{\includegraphics[width=0.8\linewidth,height=0.32\textheight,keepaspectratio]{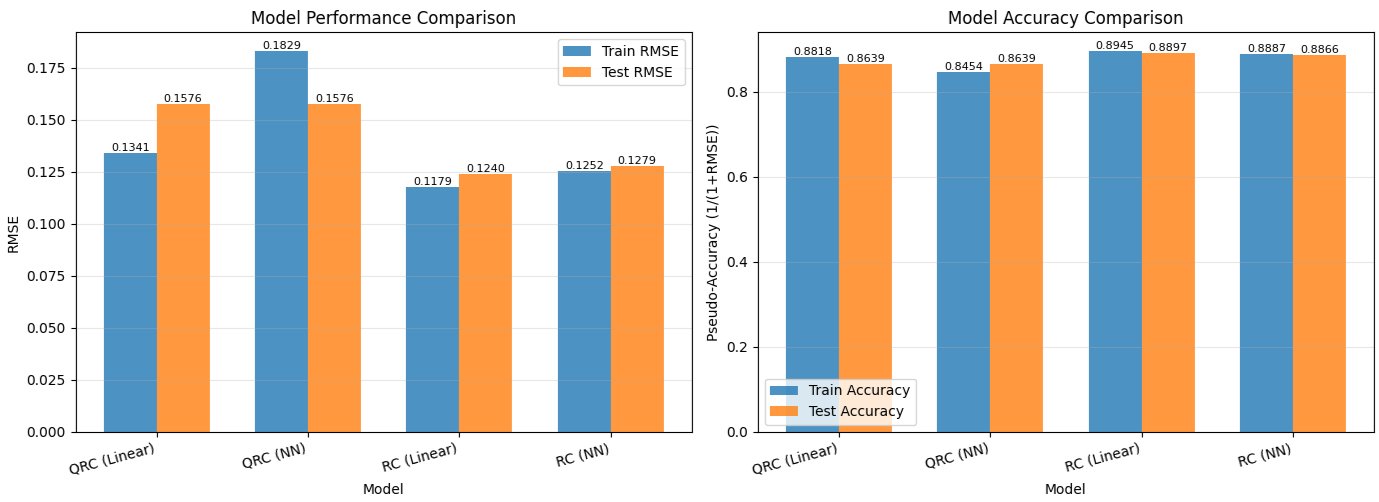}\label{uni-rc-alt }}\\[0.3cm]
    \subfloat[Multivariate losses and accuracies]{\includegraphics[width=0.8\linewidth,height=0.32\textheight,keepaspectratio]{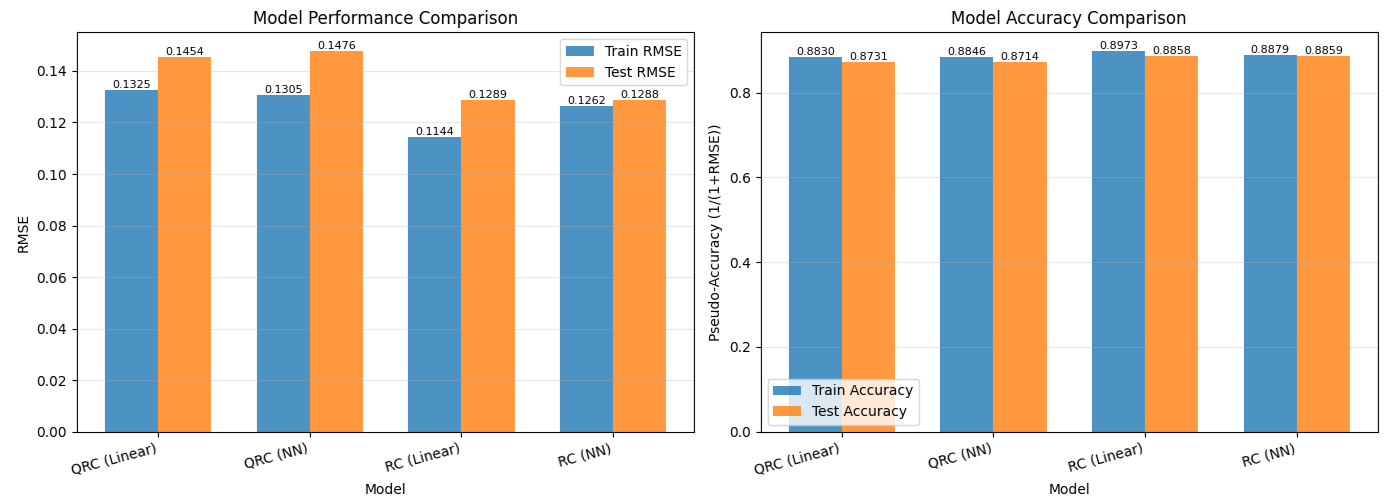}\label{multi-rc-alt}}
    \caption{Comparison of QRC and RC variant tasks.}
    \label{fig:qrc_2x2}
\end{figure*}

In the multivariate setting, where multiple correlated input channels are injected into the reservoir, QRC consistently achieves lower training and testing RMSE than the classical RC with readouts of comparable capacity Fig.~\ref{multi-rc-alt}. This improvement directly translates into higher pseudo‑accuracy values, defined as \(1 / (1 + \text{RMSE})\), for both the linear (LR) and neural network (NN) readouts. The performance gap is evident across all QRC variants, indicating that the advantage is not limited to a specific choice of readout architecture but is rooted in the underlying reservoir dynamics.

\begin{figure*}
    \centering
    \subfloat[Multivariate QRC]{%
        \begin{minipage}{0.46\textwidth}
            \centering
            \includegraphics[width=\linewidth]{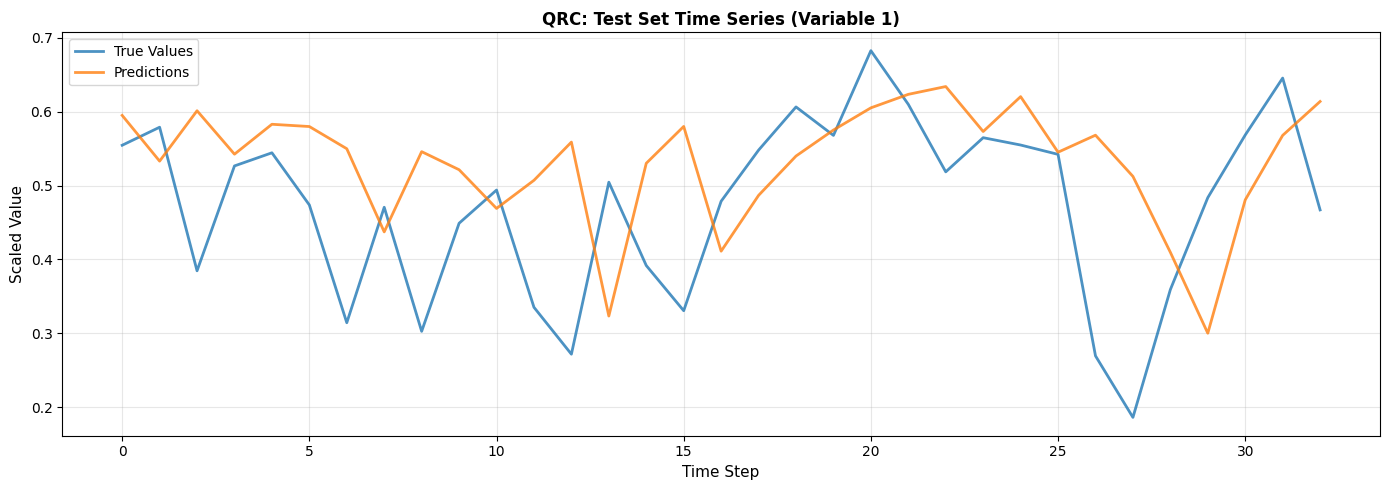}\\[0.2cm]
            \includegraphics[width=\linewidth]{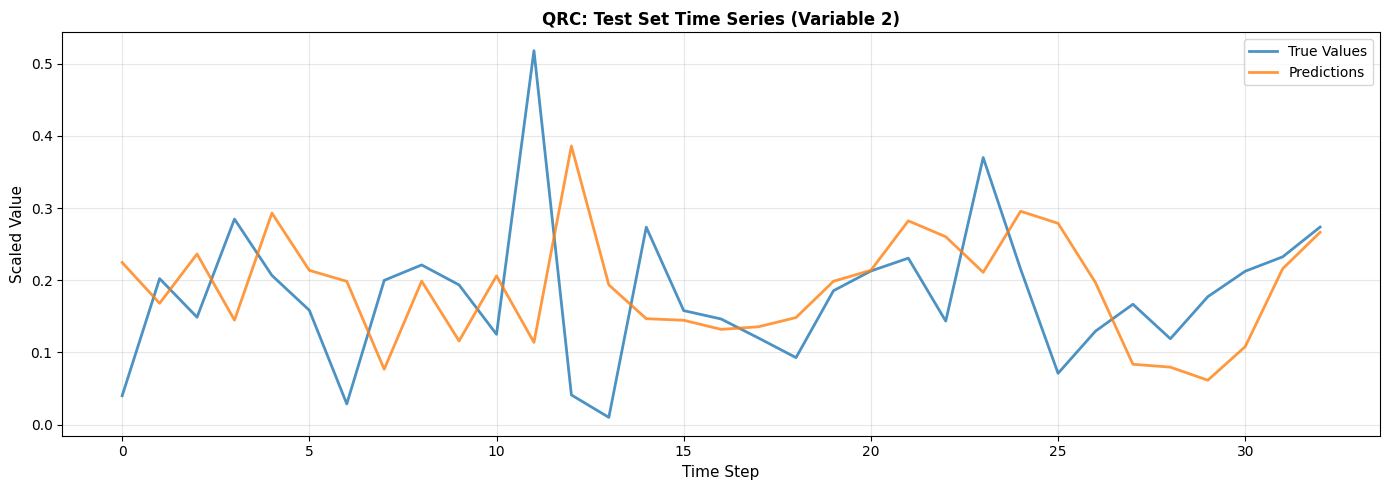}
        \end{minipage}%
        \label{fig:qrc_m_pair}
    }\hfill
    \subfloat[Multivariate NN QRC]{%
        \begin{minipage}{0.46\textwidth}
            \centering
            \includegraphics[width=\linewidth]{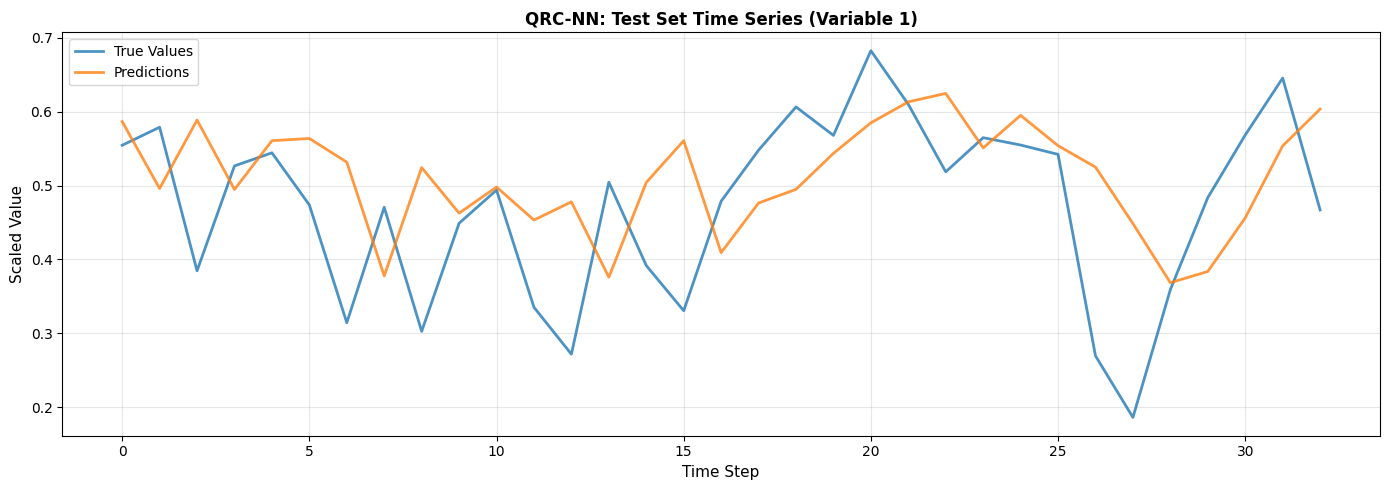}\\[0.2cm]
            \includegraphics[width=\linewidth]{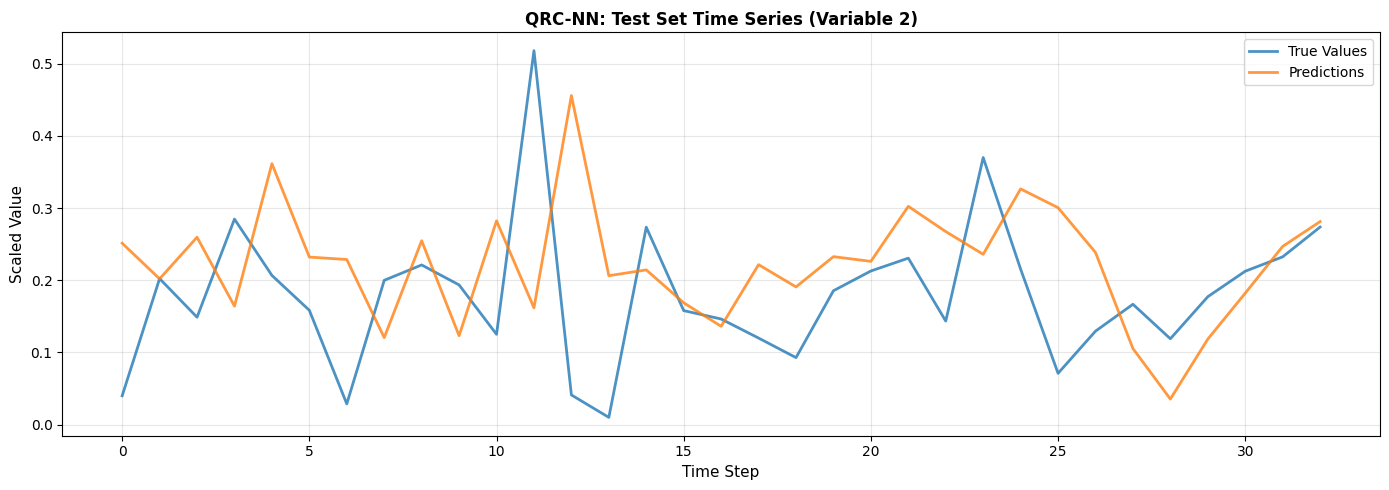}
        \end{minipage}%
        \label{fig:qrc_m_nn_pair}
    }

    \subfloat[Multivariate RC]{%
        \begin{minipage}{0.46\textwidth}
            \centering
            \includegraphics[width=\linewidth]{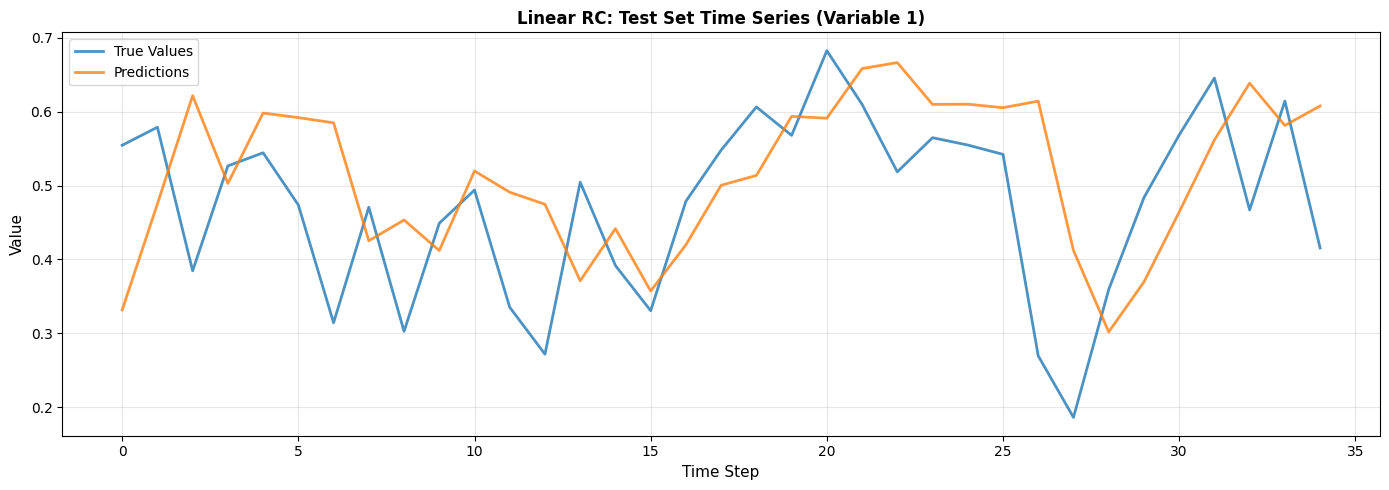}\\[0.2cm]
            \includegraphics[width=\linewidth]{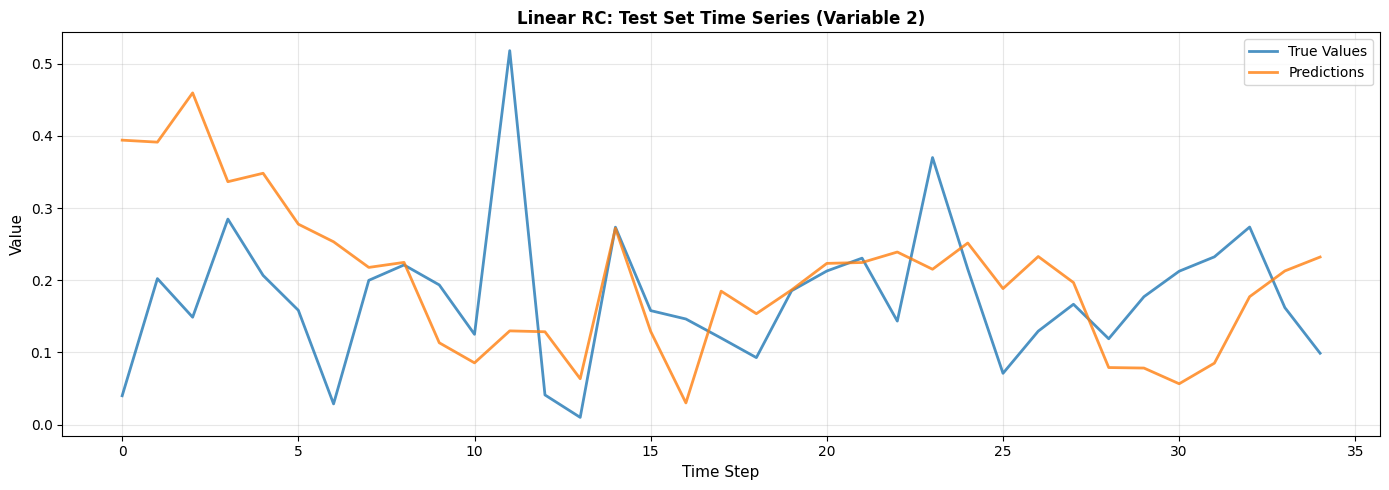}
        \end{minipage}%
        \label{fig:rc_lr_pair}
    }\hfill
    \subfloat[Multivariate NN RC]{%
        \begin{minipage}{0.46\textwidth}
            \centering
            \includegraphics[width=\linewidth]{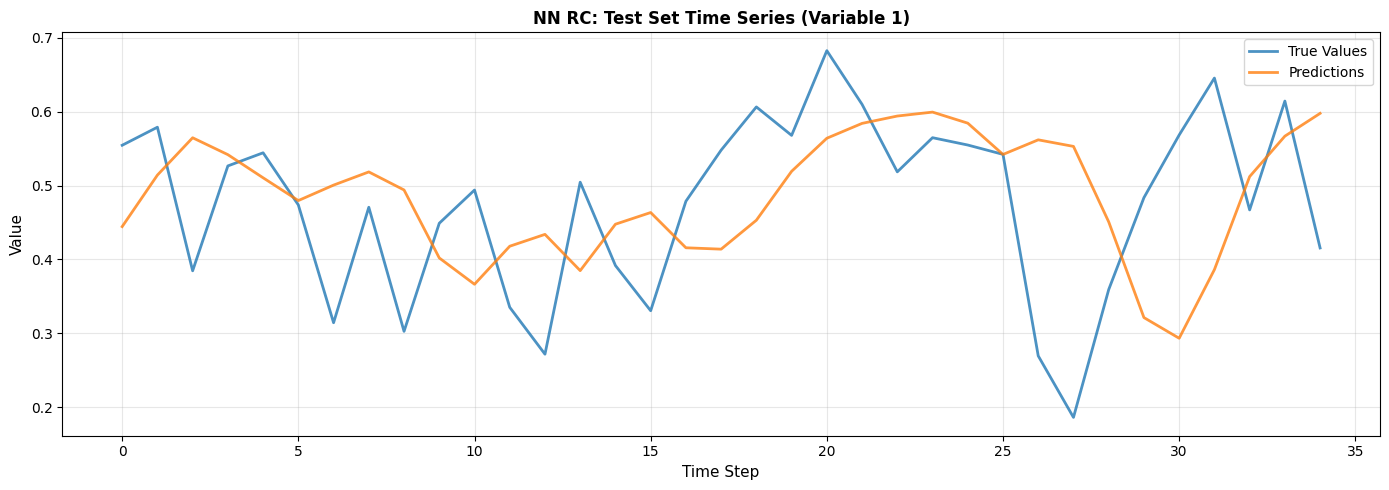}\\[0.2cm]
            \includegraphics[width=\linewidth]{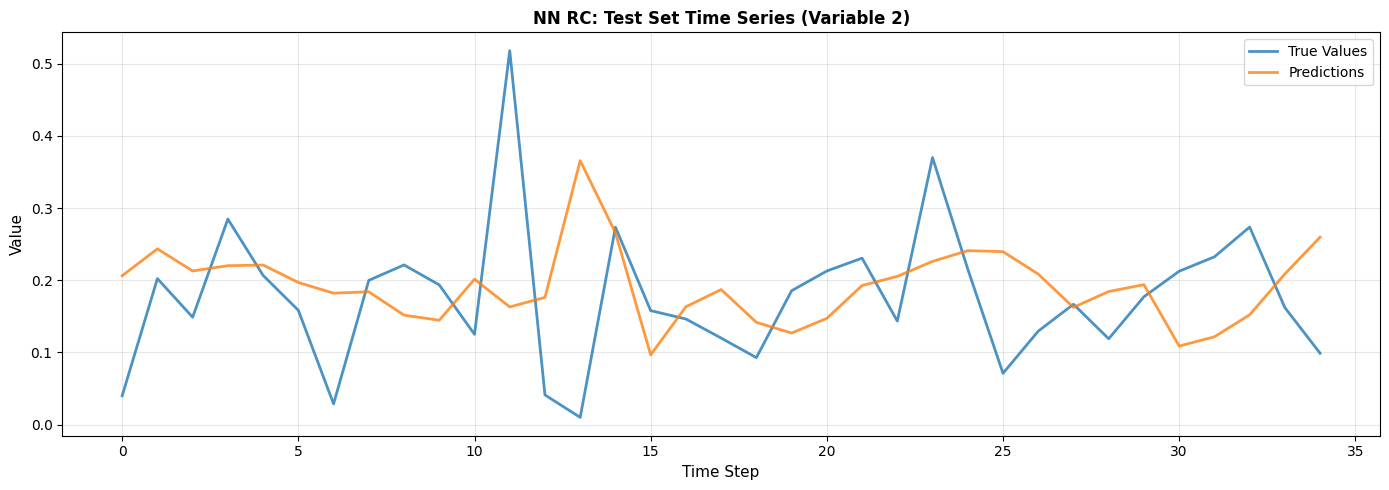}
        \end{minipage}%
        \label{fig:rc_nn_pair}
    }

    \caption{QRC and RC comparison on multivariate data.}
    \label{fig:qrc_rc_multi_pairs}
\end{figure*}

A qualitative comparison of the predicted trajectories further supports this conclusion Fig.~\ref{fig:qrc_m_pair} \& \ref{fig:qrc_m_nn_pair}. The QRC and NN QRC models more faithfully reproduce cross‑channel temporal structures, including covarying peaks, troughs, and lagged interactions between variables, whereas the RC and NN RC baselines exhibit larger deviations and occasional phase mismatches. These observations suggest that the high‑dimensional quantum dynamics of the QRC reservoir provide more expressive internal representations of multivariate inputs than those generated by the classical reservoir under similar resource constraints (see Fig.~\ref{fig:rc_lr_pair} \& \ref{fig:rc_nn_pair}). As a result, the QRC is better able to capture and exploit cross channel dependencies, leading to more accurate multivariate forecasts.

% This indicates that, for time series with relatively simple dynamics, the nonlinear mixing afforded by a reservoir already suffices, and the additional structure provided by a quantum reservoir does not translate into a substantial performance gain.

% In the multiple correlated input channels are injected into the quantum reservoir, QRC consistently yields lower train and test RMSE than an RC with an equivalent capacity readout, and this translates into higher pseudo accuracy 1/(1+RMSE)1/(1+RMSE) for both the LR and NN readouts. The quantum reservoir is able to reconstruct cross channel patterns more accurately, indicating that its highdimensional quantum dynamics provide more expressive internal representations of multivariate inputs than the classical reservoir can achieve under similar resource constraints see fig \ref{fig:qrc_rc_multi_pairs}.
% In the multivariate framework, however, the advantages of QRC become more evident. When multiple correlated input channels are injected into the quantum reservoir, QRC consistently yields lower prediction error than an RC with an equivalent capacity readout. The quantum reservoir is able to reconstruct cross channel patterns more accurately, indicating that its high dimensional quantum dynamics provide more expressive internal representations of multivariate inputs than the classical reservoir can achieve under similar resource constraints (see Fig. 6 and Fig. 7).

% Overall, our results show that for univariate prediction tasks, quantum recurrent models (QLSTM, QRC) perform comparably to their classical counterparts (LSTM, RC), with only minor improvements in error metrics. In this regime, where temporal dependencies and input structure are relatively simple, the use of quantum circuits is not strictly necessary given present‑day hardware noise and overhead.

% In contrast, for multivariate prediction tasks, the benefits of quantum models become significant. VQCs within the QLSTM provide a richer representation space that more efficiently captures inter‑channel correlations than an equivalent‑sized classical LSTM. Similarly, QRC exploits high‑dimensional quantum dynamics to encode complex nonlinear relationships among variables that are difficult to model with resource‑constrained classical reservoirs. In both cases, the improvements cannot be attributed to larger model size, as the number of trainable parameters is carefully matched.

% Despite these advantages, several practical constraints remain. Quantum models incur additional overhead due to circuit execution, and are sensitive to hardware noise and shot noise, especially for low‑dimensional tasks or applications requiring very low latency. Nevertheless, the performance observed at small scale suggests that, as quantum hardware matures and larger quantum sequence models become feasible, the advantages we observe in multivariate settings are likely to become more pronounced.

Overall, these experiments indicate that quantum recurrent models (QLSTM, QRC) deliver performance that is comparable to LSTM and classical RC in low dimensional univariate forecasting tasks, with only minor gains under matched parameter budgets. In contrast, for multivariate forecasting with multiple correlated input channels, quantum models achieve consistent but moderate improvements in error metrics and pattern preservation relative to equally sized classical baselines. These results do not demonstrate a strong quantum advantage in an asymptotic sense, but they suggest that, within current hardware and qubit constraints, quantum state spaces can be exploited as expressive feature maps that become beneficial precisely in regimes with richer cross sectional structure.

\section{Conclusion}
In this study, we compared quantum and classical sequence models for time‑series prediction, specifically QLSTM and QRC against LSTM and classical RC, in both univariate and multivariate settings on real financial data. All models were designed with closely matched numbers of trainable parameters so that differences in performance primarily reflect representational capacity rather than model size.

Our empirical results show that, in univariate tasks with relatively simple temporal structure, quantum models offer only minor accuracy gains: QLSTM and QRC achieve error levels that are essentially comparable to those of LSTM and RC, making the additional implementation overhead and noise sensitivity of current quantum hardware difficult to justify in this regime. In multivariate settings with multiple correlated input channels, however, quantum‑enhanced architectures can modestly outperform their classical counterparts, achieving lower forecasting errors and better preservation of cross channel patterns under the same parameter budget.

These findings suggest that quantum recurrent models are most promising as expressive feature mappers for high‑dimensional, strongly correlated financial time series, rather than as universal replacements for classical models. Future work will scale the architectures to larger quantum devices, explore alternative encoding and ansatz designs, and evaluate hybrid quantum–classical schemes and hardware‑aware training strategies, with the goal of further clarifying when and to what extent such quantum models provide practically relevant benefits over strong classical baselines.

\section*{Acknowledgment}
This work was supported by the German Federal Ministry of Research, Technology and Space within the funding program "Application orientied quantum computing" under Contract No. 13N17159. 

% \enlargethispage{4\baselineskip}

\end{document}